\documentclass{aastex631}

\usepackage[normalem]{ulem}
\shorttitle{AASTeX v6.3.1 Sample article}
\shortauthors{Gonçalves et al.}
\graphicspath{{./}{figures/}}

\begin{document}

\title{{\tt{CLOWN}}: The PASO cloud detection for optimization of automatic optical surveys.}

\author[0000-0002-0419-731X]{Luís Gonçalves}
\affiliation{CFisUC, Departamento de F\'{\i}sica, Universidade de Coimbra, 3004-516 Coimbra, Portugal}

\author[/0000-0002-7490-1990]{Bruno Coelho}
\affiliation{CFisUC, Departamento de F\'{\i}sica, Universidade de Coimbra, 3004-516 Coimbra, Portugal}
\affiliation{ATLAR Innovation, Ed. Multiusos, Rua Rangel de Lima, 3320-229 Pampilhosa da Serra, Portugal}

\author[0000-0002-5191-7826]{Domingos Barbosa}
\affiliation{Instituto de Telecomunica\c{c}\~oes, Universidade de Aveiro, Campus Universit\'ario de Santiago, 3810-193 Aveiro, Portugal}

\author[0000-0002-5398-9567]{Miguel Bergano}
\affiliation{ATLAR Innovation, Ed. Multiusos, Rua Rangel de Lima, 3320-229 Pampilhosa da Serra, Portugal}
\affiliation{CICGE, Faculdade de Ci\^encias da Universidade do Porto, 4169-007 Porto, Portugal}

\author[0000-0001-7412-6704]{Vitor Bonifácio}
\affiliation{Departamento de Física, Universidade de Aveiro, Research Centre on Didactics and Technology in the Education of Trainers (CIDTFF), Campus Universitário de Santiago, 3810-193 Aveiro, Portugal}

\author[0000-0001-7855-1919]{Dalmiro Maia}
\affiliation{CICGE, Faculdade de Ci\^encias da Universidade do Porto, 4169-007 Porto, Portugal}








\begin{abstract}

Orbiting space objects have become in the last decade a major nuisance impacting ground astronomy and orbiting space assets, from observatories to satellites and space stations. In particular with the rise of the satellite population in Low Earth Orbits (LEOs), space objects are becoming an even bigger threat and a strong problem to astronomical observations. To tackle these threats several coordinated surveillance networks composed of dedicated sensors (telescopes, radars and laser ranging facilities) track and survey space objects, from debris to active satellites. As part of the European Space Surveillance \& Tracking (EU-SST) network, Portugal is developing the PAmpilhosa da Serra Space Observatory (PASO), with both radio and optical telescopes dedicated to the Space Situational Awareness (SSA) domain, deployed at a Dark Sky destination. To optimize telescope survey time, we developed {\tt{CLOWN}} (CLOud Watcher at Night), an application interface that automatically monitors clouds in real time. This software can correctly trace clouds positions in the sky, provides accurate pointing information to the observation planning of the optical telescope to avoid cloudy areas. {\tt{CLOWN}} only requires the use of an all-sky camera, which is already a norm in observatories with optical telescopes and can be used with any camera, including those for which no information about its model specification do exist. {\tt{CLOWN}} does not require great computing power and it does not require the installation of additional equipment. {\tt{CLOWN}} results are very promising and confirm that the app can correctly identify clouds in a variety of different conditions and cloud types.

\end{abstract}



\section{Introduction} \label{sec:intro}

The last few decades have seen an explosion in space exploration, with a rapid increase in the number of satellite launches, with several private companies creating their mega-constellations with an expected number topping more than $\sim42000$ satellites \citep{EstudoLEO}. As we write, there are over 6000 satellites including more than 5000 constellations satellites in orbit \citep{leonumbers2023, linkNacoesUnidas,linksatelites}, and these numbers are expected to dramatically increase in the next decade. The continuous accumulation of satellites and associated space debris in orbit over time is today a major concern \citep{MCNALLY1999255,Rossi}. Even small debris can greatly damage an active satellite due to their high speeds ($\sim28000$ km/h). With the rising number of objects the risk of collisions increases which will in turn create even more debris, creating a cascade effect known as ``Kessler Syndrome"  \citep{Kessler1978CollisionFO} that may deeply affect the orbital space around the Planet.

Of particular concern to astronomy, planned LEO satellite mega-constellations and the associated space debris field reflect light from the sun and are visible from optical to infrared and their beacons and passive Radio frequency reflections from ground emitters are detected by ground detectors as well. This orbital presence constitutes a threat to the dark and quiet skies and will have negative impacts on astronomy with wider societal impacts. These concerns have motivated the creation of the International Astronomical Union (IAU) Centre for the Protection of the Dark and Quiet Sky from Satellite Constellation Interference (IAU CPS) \citep{2023arXiv231102184B} to address the effects of this ever-growing satellite population and propose mitigation measures and reduce impacts on astronomical surveys.

As advocated by the IAU CPS\footnote{https://cps.iau.org/}, the dark and quiet sky is an essential requirement to facilitate fundamental research and technology development, programmatic observations for key priorities included in the SSA domain like planetary defence, and is essential for basic services like space navigation and geolocation. As a major example of the current impacts of mega-constellations on astronomical facilities, the Vera C. Rubin Observatory, a large optical telescope with a wide field of view scheduled to commence operations in 2024 whose main task will be the production of the world's largest synoptic astronomical survey, the Legacy Survey of Space and Time \citep{2019ApJ...873..111I} that will have up to 30\% of its images at the beginning and end of each night \citep{2021Msngr.184....3W}, the night moments when LEOs are brighter, impacted by one or more satellite streaks from the almost 42,000 planned SpaceX satellite constellation.

Indeed, as shown above,  LEO constellations may compromise the effectiveness of astronomical infrastructure and will likely impact national and international investments in astronomy. Adding to this, the rise of space debris in particular in LEO orbits poses a direct threat to space assets.

This enormous and rising number of space objects has led to the development and world deployment of automated networks to detect these debris, verify orbital occupancy, improve space object orbital parameter determination, and provide alert services. This should evolve soon towards a Space Traffic Management system that may include collision avoidance, fragmentation detection, and Earth reentrance services and database management. Major space-faring nations developed their own SSA/SST programs, notably the US with the Spacetrack\footnote{https://www.space-track.org/auth/login} network \citep{inproceedings, bulmann}, China \citep{DU20178}, India \citep{PRASAD2005243, ADIMURTHY2006168}, Russia (as an update and follow-up of the previous Soviet network), Brazil \citep{froehlich2020space,2012cosp...39.1381N} and more recently Europe with the European Space Surveillance and Tracking\footnote{https://www.eusst.eu}  (EU-SST). These expanding SSA ground sensor networks are proof of the great investments in this area  \citep{Tingay_2013, Jiang_2022} and besides their regional priorities they constitute an important source of relevant information on objects affecting astronomical observations.

\subsection{The PASO optical telescopes} \label{sec:paso}

As part of the Space Surveillance \& Tracking - Portugal (SST-PT), a node to the EU-SST, the Pampilhosa da Serra Space Observatory (PASO) hosts the rAdio TeLescope pAmpilhosa Serra (ATLAS), a monostatic radar tracking sensor currently in test phase \citep{signals2010011}, a wide field of view telescope for GEO space surveillance operated under the auspices of the Portuguese MoD (PT-MoD) and a double telescope system \citep{coelho2021new}, installed in the Summer of 2022, and currently in integration. PASO is developed within the Dark Sky "Aldeias do Xisto" Tourism destination, a dark sky area certified by the Starlight Foundation and thus guarantees low levels of light pollution and good sky background ($\ge $21.1 mag/arcsec$^2$) and at least 50\% of clear nights per year. The combined operation of radar and optical sensors takes advantage of radar measurements which give precise radial velocity and distance to the objects, while the telescope gives better sky coordinates measurements. With the installation of radar and optical sensors, PASO can extend the observation time of space mega-constellations and space debris and correlate information from optical and radar provenances in real-time. During twilight periods both sensors can be used simultaneously to rapidly compute new orbits for LEO objects, eliminating the time delays involved in data exchange between sites in a large SST network. This concept will not replace the need for an SSA/SST network with sensors in multiple locations around the globe but will provide a more complete set of measurements from a given object passage, and therefore increase the added value for initial orbit determination, or monitoring of reentry campaigns at a given location. PASO will contribute to the development of new solutions to better characterize the objects improving the overall SST capabilities and constitute a perfect site for the development and testing of new radar and optical data fusion algorithms and techniques for space debris monitoring.

The PASO double telescope system can observe a maximum FoV of 4.3$^\circ\times2.3^\circ$.  It is equipped with  B, V, R, I, H$\alpha$ and [OIII] filters that allow the acquisition of light curves in different wavelengths (up to two simultaneously). Its main goal will be to automatically track objects in Low Earth Orbits (LEO) in particular at dawn and before sunset but it will also have the capability to track and survey objects in Medium Earth Orbit (MEO) and Geosynchronous Orbit (GEO) when not observing LEOs. Besides its SSA operational duties, the WFOV twin telescope will be used for transient science or to monitor variable sources in the sky like Andromeda Novae \citep{2019Natur.565..460D,2022ApJ...939....6A,2022yCat..22570049C}. The system is being prepared to operate with minimum to none human intervention while it maximizes observational time and optimizes the data processing pipeline procedures.

\begin{table}[h!]
\label{table_paso}
  \begin{center}
  \caption{PASO optical SSA instruments. }
  \label{tab2}
 {\scriptsize
  \begin{tabular}{|l|c|c|c|c|c|c| }\hline 
{\bf Optical Instruments}      &  {\bf Bands}   & {\bf Size}   &  {\bf Mount}   &  {\bf Type} & {\bf Speed} & {\bf FoV} \\
                &           &   &      &   &  & \\ \hline

Twin WFOV Telescope{*}  &   BVRI, H$\alpha$    &       $2\times0.3$ m             &  Equ & Riccardi-Honders &  $\le 40^\circ$/sec & $\sim4.3^\circ\times2.3^\circ$  \\
  (LEO - GEO)              &      [OIII]     &   &      &  &  &  \\ 
WFOV Optical Telescope\textsuperscript{\textdagger}  & --                      & $0.35$ m       &  Equ  & RASA 14"& $\le 6^\circ$/sec & $\sim 2.6^\circ\times2.6^\circ$ \\ 
      (MEO/GEO)          &           &   &      &   & &  \\ \hline


  \end{tabular}
}
 \end{center}
\vspace{1mm}
 \scriptsize{
 {\it Notes:}\\
  {*}A PT-MoD, Instituto de Telecomunicações and University of Coimbra partnership.\\
  \textsuperscript{\textdagger}Telescope property of PT-MoD, operating within the EU-SST network.}
\end{table}

An SST observational planning will consider the position and phase of the moon, the position of the Milky Way, the shadow of the earth, and meteorological conditions to improve the quality of data and protect the equipment from damage. Cloudy weather impacts the availability and efficiency of the optical telescopes when used automatically by compromising the captured data. LEO satellites are only visible for a fraction of the night (at dawn and dusk essentially), and are blocked for most of the night by the Earth's shadow.  The optical telescope operation needs to be optimized to avoid cloudy patches and avoid losing any observation and image processing time. To maximize the system efficiency and availability, we designed an automatic and real-time cloud detection system to support the operation of the optical telescopes. The proposed solution aims to keep costs down, making use of equipment (an all-sky camera) already present in any typical observatory. This solution is easy to deploy in any observatory with telescopes preforming automated observation, either the ones operating for SSA or the ones dedicated to astronomical duties (eg. DES \cite{DES}; Vera C. Rubin LSST \cite{LSST}; Pan-STARRS \cite{Pan-STARRS};  J-PAS, J-PLUS, S-PLUS \cite{JPAS}, \cite{JPlus}, \cite{SPLUS}).

\section{{\tt {Clown}}: CLOud Watcher at Night} \label{sec:style}

\begin{figure}[!ht]
    \centering
    \includegraphics[width=120mm,scale=1]{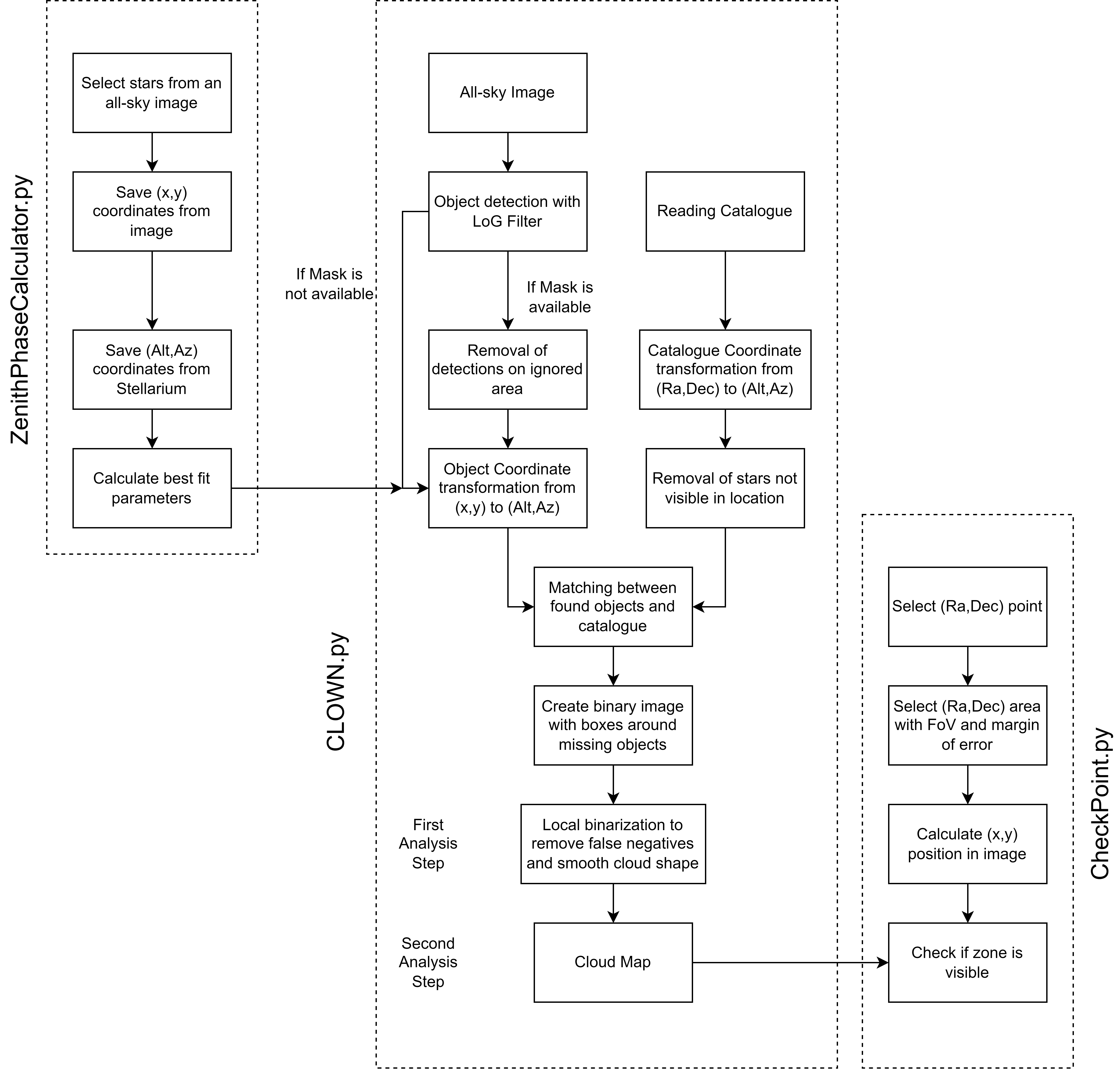}
    \caption{Basic flow diagram of the {\tt{CLOWN}} pipeline.}
    \label{fig:flowchart}
\end{figure}

{\tt {CLOWN}}: CLOud Watcher at Night, is a software developed to detect clouds obscuration on all-sky images, and optimize astronomical or space surveys. Although it was developed to increase survey time for SSA purposes it can be used as an ancillary tool to optimize astronomical large sky area surveys and streamline telescope pointing decisions. Apart from increasing the availability of optical surveys, it can also be used for other goals such as study the presence of clouds in the region either for meteorological reasons or to analyze the applicability of that location for an optical telescope. The provided in real time cloud maps may also constitute a valuable tool for the operation of optical communication ground stations. For instance the number of cloudless nights is an important parameter for Dark Sky astrotourism certifications. The complete operation flowchart of the software is presented in figure \ref{fig:flowchart}. The algorithm was developed using the programming language Python and it is available on \href{https://github.com/LuisGoncalves1/CLOWN}{GitHub}\footnote{https://github.com/LuisGoncalves1/CLOWN} and on Zenodo \citep{CLOWNZenodo}.

The automatic cloud detection pipeline has to deal with the fact that clouds may be brighter or darker than the apparent sky background. The brighter clouds may result from ground artificial light reflections or produced by the dispersion of light emitted by a bright sky source like the Moon; hence they will be referred as visible. On the other hand, the clouds darker than the sky background are referred here as invisible. Cloud detection is based on the assumption of the presence or absence of stars in the image since their  positions are well known from star catalogs. This enables the detection of both visible and invisible clouds as in both cases they will block the stars from appearing in the image.


\subsection{All-sky images}

An All-Sky camera observes the entire sky above it via a fish-eye lens and takes images with a large FoV up to $2\pi$ steradians. There are different types of the fish-eye lens using different projections (or curved deformation). The image fits into the sensor with usually four main projections: stereographical, equidistant, equisolid angle, and orthographic. Although each projection has its own special properties and applications the equidistant projection is the most suited for our purposes since it creates a linear relation between the distance to the center of the image and the corresponding angular coordinates \citep{lentes}. 

The camera available at PASO is the Occulus All-Sky Camera 180$^\circ$ \citep{camera}, with camera information in Table \ref{tab:Occulusparameter}. Its sensor has a 1392$\times$1040 pixel resolution, with a pixel size of 4.65 $\mu$m and a fish-eye lens of focal length f = 1.55 mm enabling an equidistant image projection. The camera is equipped with a policarbonate dome to protect the lens from the environment and an heating system to avoid condensation. These features make it ideal for sky monitoring and provide ancillary pointing information to the telescope operators during the entire night. The images have 30 seconds of time exposure, enabling the observation of Hipparcos catalog \cite{Hipparcos} stars up to 6.5 magnitude across the entire FoV. For our PASO FoV, this results in over 2000 stars on average per image. This is enough to analyze the entire sky hemisphere with enough precision for the operation of the WFOV double telescope system at PASO. 

\begin{table}[h!]
\caption{All-sky cameras parameters (\cite{camera})}
\centerline{
\begin{tabular}{|c|c|c|c|c|}
\hline
Camera Model & Resolution & Pixel size & Focal length & Projection   \\ \hline
Occulus Camera    & 1392x1040  & 4.65 $\mu$m    & 1.55mm       & Equidistant              \\ \hline
OPD Camera  (Brasil)   & 640x480  & Unknown    & Unknown       & Unknown              \\ \hline
\end{tabular}}
\label{tab:Occulusparameter}
\end{table}

 We use \emph{AllSkEye} \citep{AllSkEye}, a freely available app, to capture the all-sky images.  Although the camera is shipped with its custom-made photo acquisition software, it still needs to be reset daily and may not comply with the high availability requirements set by an SSA or astronomical network. For this reason, the camera shipped software was not considered the best choice for an automated system requiring a longer operation time. \emph{AllSkEye} suite provides a large integrated set of options that facilitate this goal:  i) it allows for automatic photo capture between sunset and sunrise with a controlled time for each photo; ii) it has an inbuilt protection decision loop against high-intensity illumination to avoid damaging the camera's sensor; iii) it enables the use of a horizon mask to remove constant obstacles in the camera field of view including uneven horizon masks for tree lines, tall buildings, etc. The output file can be saved in both FITS or {\tt jpeg} files. Most importantly, this software also has the major benefit of being agnostic to most camera models and thus can be used in any observatory with an already custom-installed camera.

Most observatories with optical telescopes have all-sky cameras but their characteristics may be unknown if they are older models. To confirm the potential of {\tt {CLOWN}} with a generic all-sky camera, we also tested the software using images from the All-sky Camera II of the "Observat\'orio do Pico dos Dias" (OPD) in Bras\'opolis, Brazil, on which we have no technical information on the camera model specifications . These images are made publicly available by the Laboratório Nacional de Astrofísica at \citep{LNA}, with a refreshing rate of one image/minute. This camera has a 640$\times$480 pixel resolution, and it captures images with a 60-second exposure time.

\subsection{Star Detection}

The first step is to perform the source extraction. Stars in an all-sky image will be bright sources or blobs spread within a dark background, creating a large contrast in their locations. Identification of blob-like structures through contrast enhancement (CE) methods of image processing include non-linear transformations \citep{Peli:90,BEGHDADI1989162,Gonzalez}, histogram-based techniques \citep{109340,Pisano,4560077}, contrast-tone optimization \citep{zhao2010linear}, frequency domain methods \citep{kingsbury1999image,Loza,7010797,Isar11}.  Among the most used methods in digital processing, contrast detection in still images has been pursued using the Sobel, Prewitt or Roberts operators \citep{7095920}. The examples given here are tools used for contrast detection that are very useful for vertical and horizontal edge detection. But additional analysis would be needed since our goal is detect the blobs centers. Although we did not pursue extensive method comparisons through the previously identified techniques, we verified the Laplacian of Gaussian or LoG method \citep{1980RSPSB.207..187M} is also used for contrast detection but it is also sensitive to blob-like structures and that is why it is the method of choice for the blob detection application, regardless of the field of study. The Laplacian is a 2-D isotropic measure of the 2nd spatial derivative of an image and therefore sensitive to regions of rapid intensity change and often used for edge detection. Natural noise variations present in every image dataset are smoothed via a Gaussian blurring step, which improves sensitivity to blob-like structures. It also creates another advantage for blob detection as the response will be sensible to the blob radius, maximum when the blob is close to $\sqrt2 \sigma$, where $\sigma$ is the standard deviation set for the Gaussian blurring step. This property allows the study of blob-like structures' size. By using the associative property of the convolution operation, these two steps can be applied simultaneously in a single convolution filter significantly decreasing the computational cost.

Our algorithm follows the \emph{blob\_log} method from the \emph{scikit-image} package \citep{scikit-image}. This applies the LoG filter to the image and outputs the position and size of the detected objects. The LoG filter accepts several parameters to optimize results; of those, there are four who are most important which are: $min_{\sigma}$, $max_{\sigma}$, $num_{\sigma}$, and the threshold. The first two parameters control the minimum and maximum $\sigma$ values used for the Gaussian filter respectively and the third controls the number of points within the $\sigma$ range. Since for our application we are not interested in analyzing in detail the size of the blobs, after some testing we reduced the value of $num_{\sigma}$ as much as possible to decrease the computational cost. 

The LoG method enables the study of the blob size, so we were able to determine the commonest $\sigma$ values in our image. We found that the smaller objects detected had a $\sigma \simeq 1$ and the largest $\sigma \simeq 1.5$, and that is why we chose those values for our variables. With these values all detections were correctly identified so we decided to keep $num_{\sigma} =2$ to decrease the computational cost. These values for $max_\sigma$ and $min_\sigma$ were confirmed after checking the LoG's property enabling the study the size of the blobs. These values may slightly change for different camera models but should be constant over time with reasonable meteorological conditions. The last important parameter is the threshold for a positive spike to be considered a detection. Choosing a very low value might detect all blobs but end up with a large number of false positives, whereas if the value is too big, the method fails to detect a large number of objects leading to false negatives. Therefore we prefer to produce a few additional false positives that can be easily removed in the posterior iterations of the algorithm, finding after testing that a threshold of 0.013 provides a good result. We let these parameters free in the configuration file to enable easier control.

\begin{figure}[!ht]
\centering
\begin{minipage}{.48\textwidth}
  \centering
  \includegraphics[height=75mm]{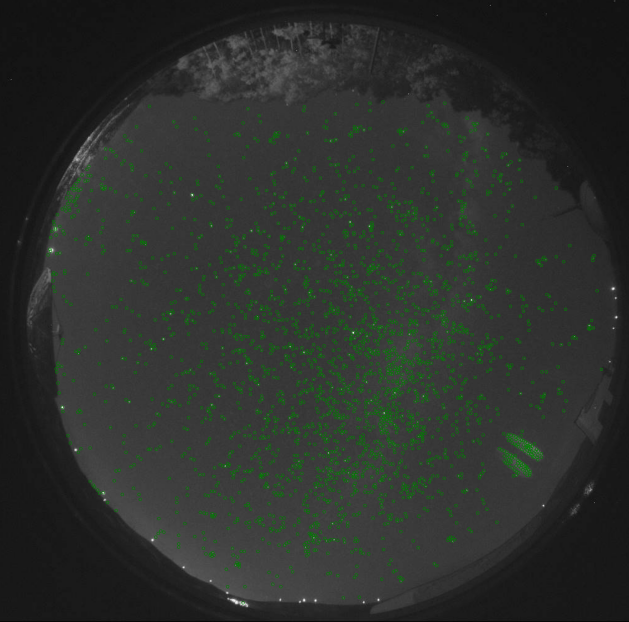}
\end{minipage}%
\begin{minipage}{.51\textwidth}
  \centering
  \includegraphics[height=75mm]{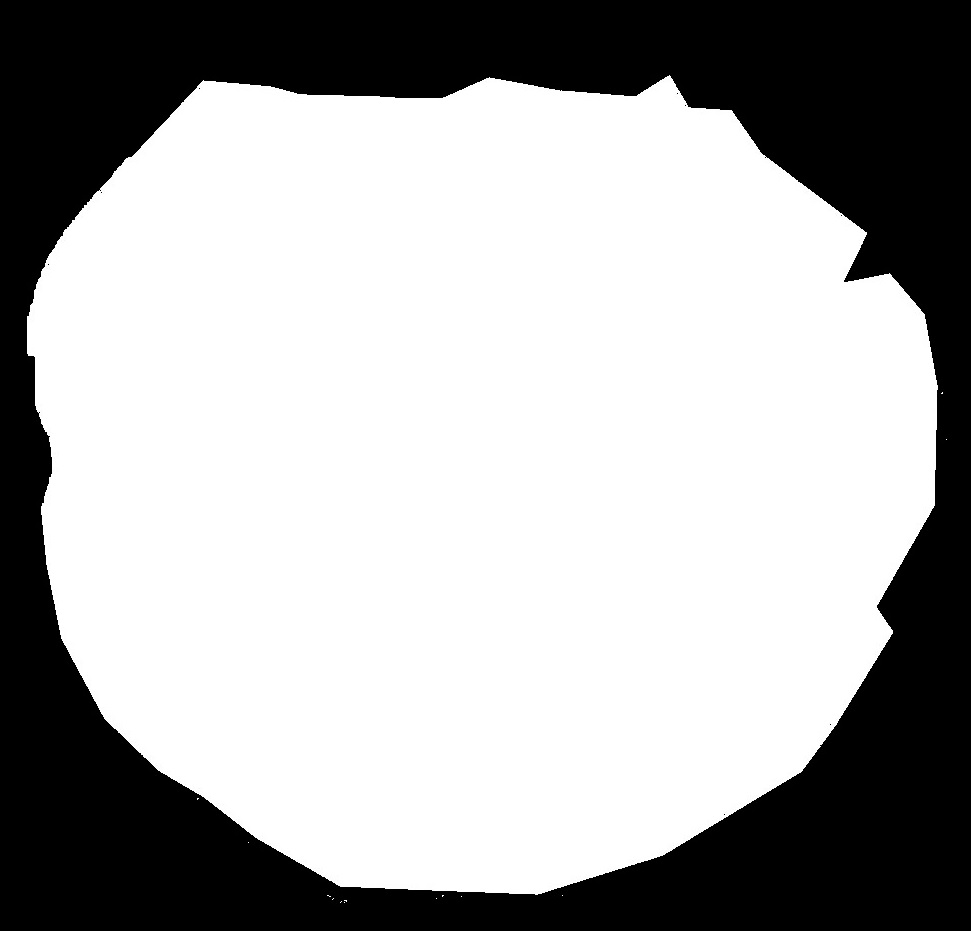}
\end{minipage}
\caption{Left image: an example of the sources detected in {\tt{CLOWN}} is displayed by green circles in an all-sky image from PASO. Right image: the mask created for the PASO all-sky camera (area of interest in white) to eliminate false detections on the horizon. A small area with several detections can be seen on the left image, which is caused by a glint caused by an illumination lamp in the area. It can be ignored because it is a small area in the sky. If the glint becomes persistent it can be resolved by including this zone in the mask or even better by physically remove that light source.}
\label{fig:detections}
\end{figure}

Figure \ref{fig:detections} (left) shows the results after the application of our LoG method to an all-sky image from PASO. The objects are correctly identified throughout the image. The presence of an uneven horizon with obstacles such as buildings or trees leads to a number of false detections which contributes to slowing down the algorithm execution. To eliminate false detections and reduce the computational cost we created a mask image of the horizon profile around the observatory in black-and-white for sharper edge detection - where white is the area of interest. With this mask, we can remove all detections in low-elevation areas that are not important to our observational operations. Besides the manual creation of a mask, a minimum sky elevation can also be specified in the configuration file as a user requirement. The mask created for the PASO all-sky camera is presented in Figure \ref{fig:detections} (right). After the first horizon determination, the resulting mask was edited with a general purpose image editing tool (such as GIMP), a relatively quick process that only needs to be done once. Codes such as the Simulating Horizon Angle Profile from Elevation Sets (SHAPES) \cite{Bassett_2021}\footnote{https://github.com/npbassett/shapes} that were developed to automatically create a mask for low radio frequency global 21-cm experiments using Google maps information were not thought to be required to our needs but could also be used at the expense of a more complex procedure and computation time.

\subsection{Star Matching}
\label{sec:StarMatching}

\begin{figure}[!ht]
    \centering
    \includegraphics[width=\textwidth]{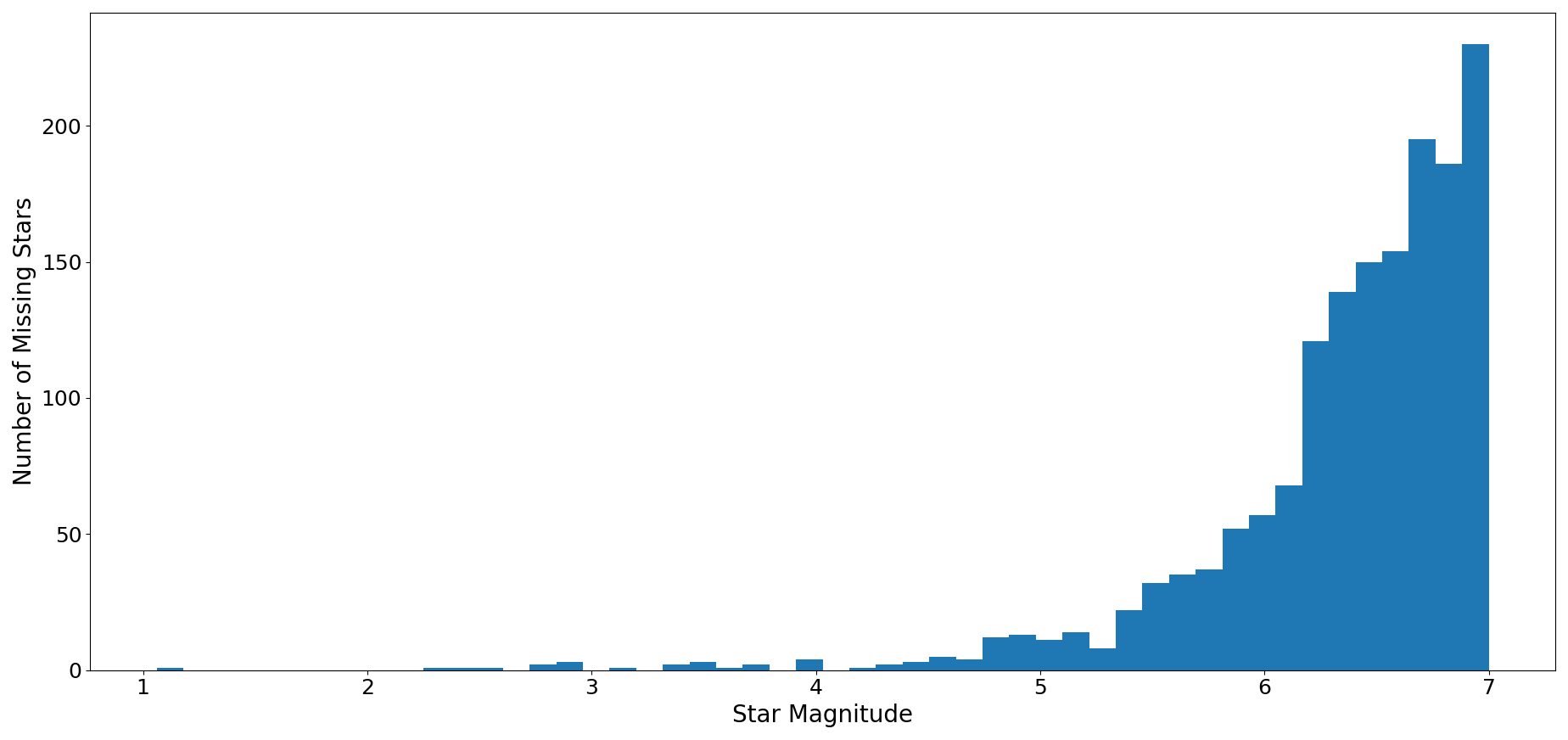}
    \caption{Relation between number of missing stars and their magnitude.}
    \label{fig:MissingStars}
\end{figure}

The algorithm's second step proceeds to a comparison of the objects found to those in a catalog such as the Hipparcos 2000 \citep{Hipparcos}. This catalog has enough objects in the range of magnitudes of interest for the application (118322 stars). For the comparison between the detected stars and the stars in the catalog, we filtered the objects in the catalog to a magnitude less than or equal to 5.5, as those were the ones that consistently showed up in our images. For higher magnitudes, the number of false negatives increases significantly as can be seen in figure \ref{fig:MissingStars}. On average, we had over 2000 stars per image. To compare the detected objects with the catalog, we used the astropy library \citep{astropy:2013,astropy:2018}. We found that in the case of a big cloud illuminated by the moon, the code treated the cloud as a very large set of small objects and did not identify any stars in the non-cloudy region. To avoid this code ill-decision we implemented a pre-condition assuming that for a very high number of detections (much bigger than expected by comparing to the catalog) it would assume that the detections were due to the clouds and instantaneously draws the clouds edges based on the detected points instead of comparing them with the catalog.

\begin{figure}[!ht]
\centering
\begin{minipage}{.48\textwidth}
  \centering
  \includegraphics[width=75mm,scale=1]{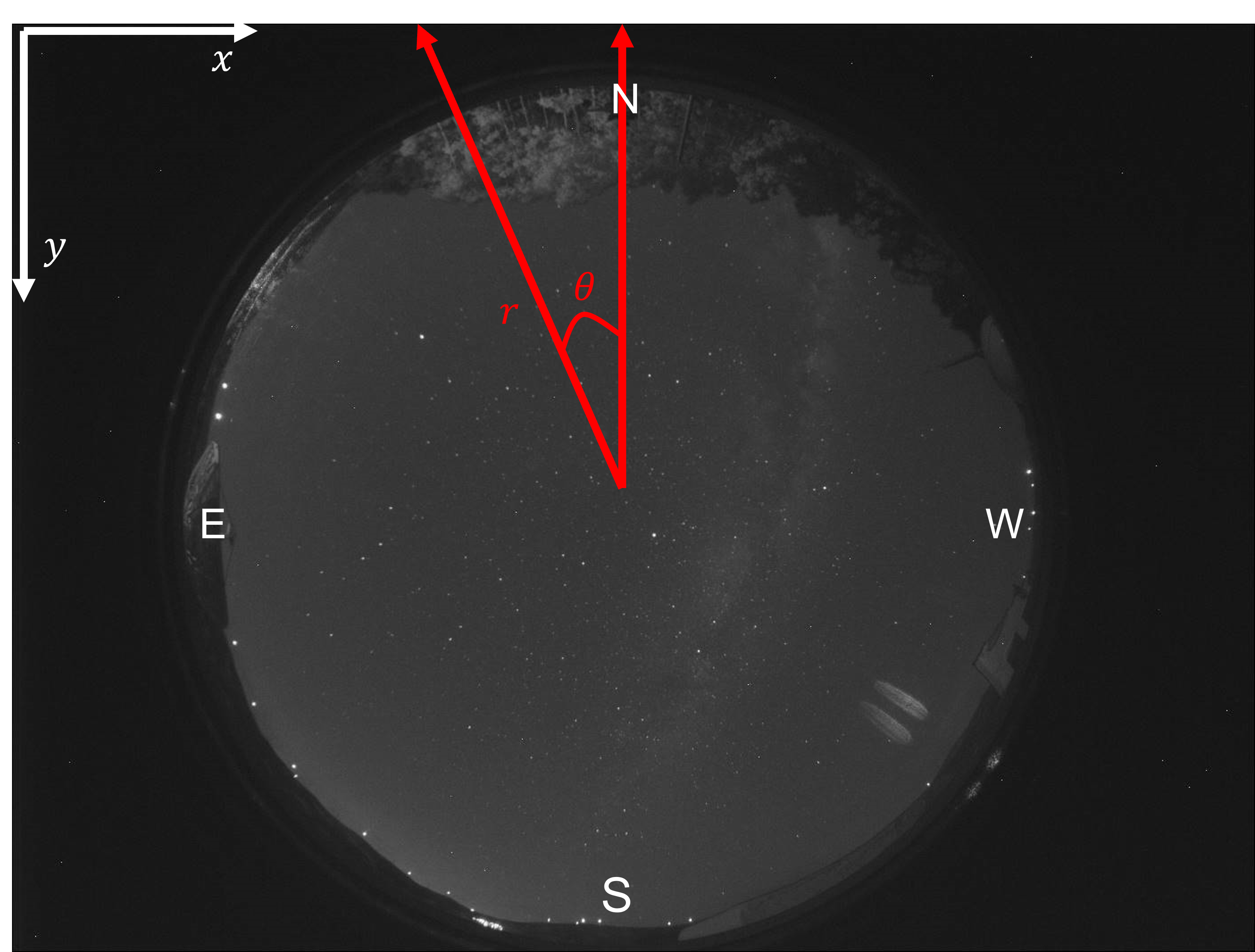}
\end{minipage}%
\begin{minipage}{.48\textwidth}
  \centering
  \includegraphics[width=75mm,scale=1]{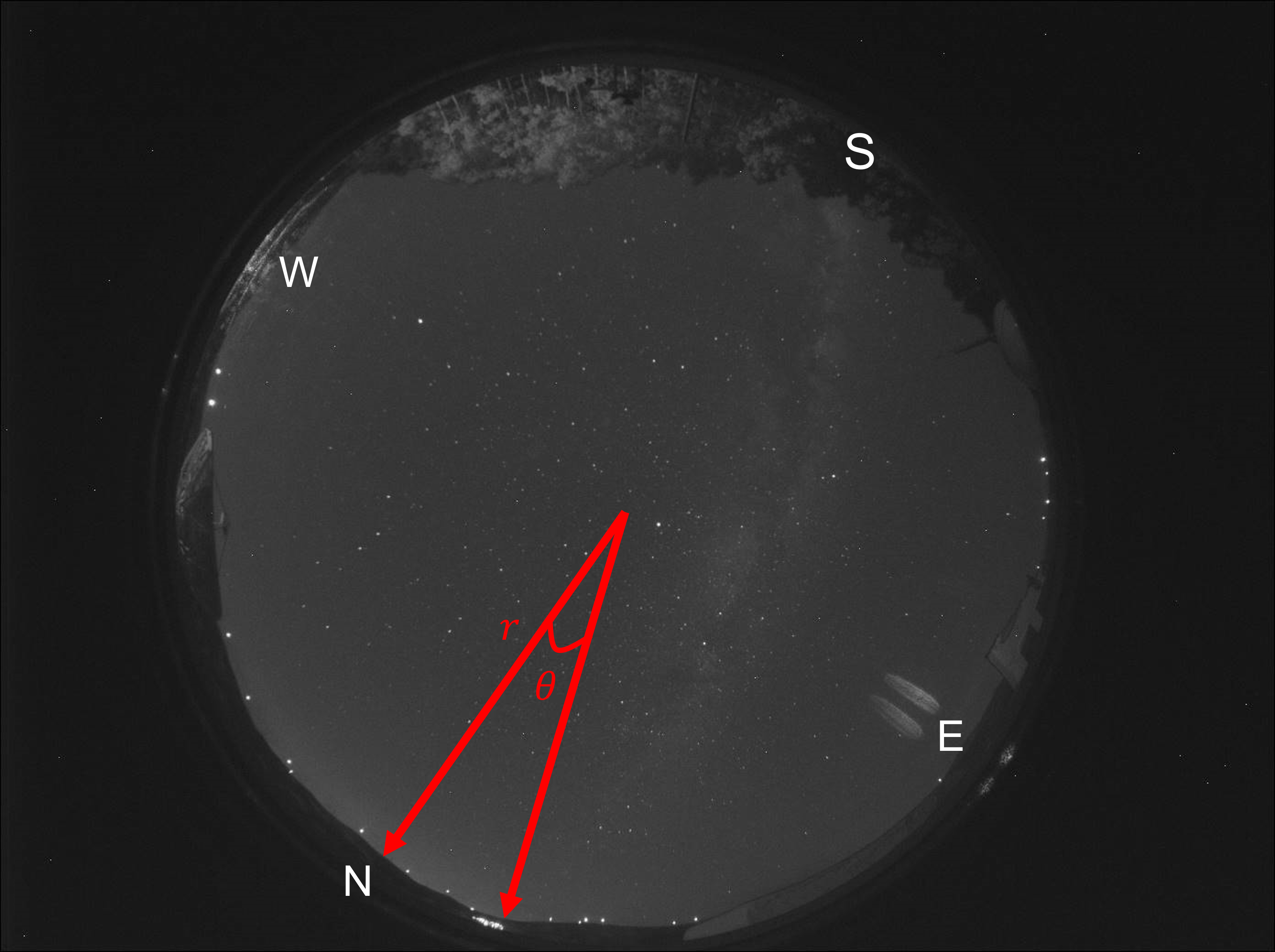}
\end{minipage}
\caption{Illustration of the polar coordinate transformation. Left image: the image pixel coordinate system (x,y) is drawn on the top left in white, and the polar coordinates system for a perfect camera installation is overlaid in red. On the right image, we have the polar coordinate system for the real system, obtained after correction for the fixed angle offset.}
\label{fig:axes}
\end{figure}

For the comparison, we transform the coordinates of our image pixels (x,y) to match the coordinates of the catalog stars (Ra,Dec), or vice-versa. The transformation from celestial coordinates (Ra,Dec) to horizontal coordinates (Alt,Az) is easily performed through the \emph{astropy} library \citep{astropy:2013,astropy:2018}. In detail, we first transform the pixel coordinates in the image into polar coordinates using the center of the image as the origin, as seen in figure \ref{fig:axes}. After checking the alignment (meaning the North is perfectly vertical in the image and the zenith is exactly the center of the image) the value of $\theta$ will be equal to the azimuth. In most cases, there will be a fixed offset that can easily be corrected. With the value of $r$ from the polar coordinates, we can use the corresponding camera transformation to obtain the corresponding angle $\phi$, which will be the zenith distance (Alt$ = 90 - \phi$).

For a perfect camera installation, the center pixel of the image should coincide with the zenith. Since in real conditions this may not be the case we need to find the zenith pixel and the related coordinate offsets with the following steps : i) the algorithm finds the brightest stars in the image, analyzes all the image across various values of $r$, and records both their (Alt,Az) and (x,y) coordinate pairs; ii)  it then compares the found coordinates to Stellarium\citep{Stellarium} database to check the (Alt,Az) coordinates of each star; iii) it  calculates the zenith pixel coordinates and the angular offset to the North pixel to minimize the error in the coordinate system transformation [(x,y) $\rightarrow$  (Alt,Az)]. 

For an application to other cameras without known model specifications (like focal distance, pixel size, and projection type), we assume by default a linear transformation (equidistant) for the image processing. In this case, it is easier to estimate the ratio between the number of pixels in the image and the corresponding angle as we know that either the width, height, or diagonal will correspond to 180$^\circ$. This transformation also has the advantage of creating an intermediary relation $\frac{r}{\theta}$ when compared with the other projections, enabling it to be applied more generally. In the case of the OPD camera we estimated that the number of pixels corresponding to the width of the image will approximately correspond to 180$^\circ$.

Using this transformation assumption we obtained a median error of 0.42$^\circ$ for the Alt/Az coordinates with the PASO camera and a 0.61$^\circ$ median error with the OPD camera. Of course, a good knowledge of the camera model properties leads to smaller errors in the determination of the star's coordinates. But we checked that even with an unknown camera model, the results are sufficiently good for our application to enable good perator decisions.

\begin{figure}[!ht]
    \centering
    \includegraphics[width=\textwidth]{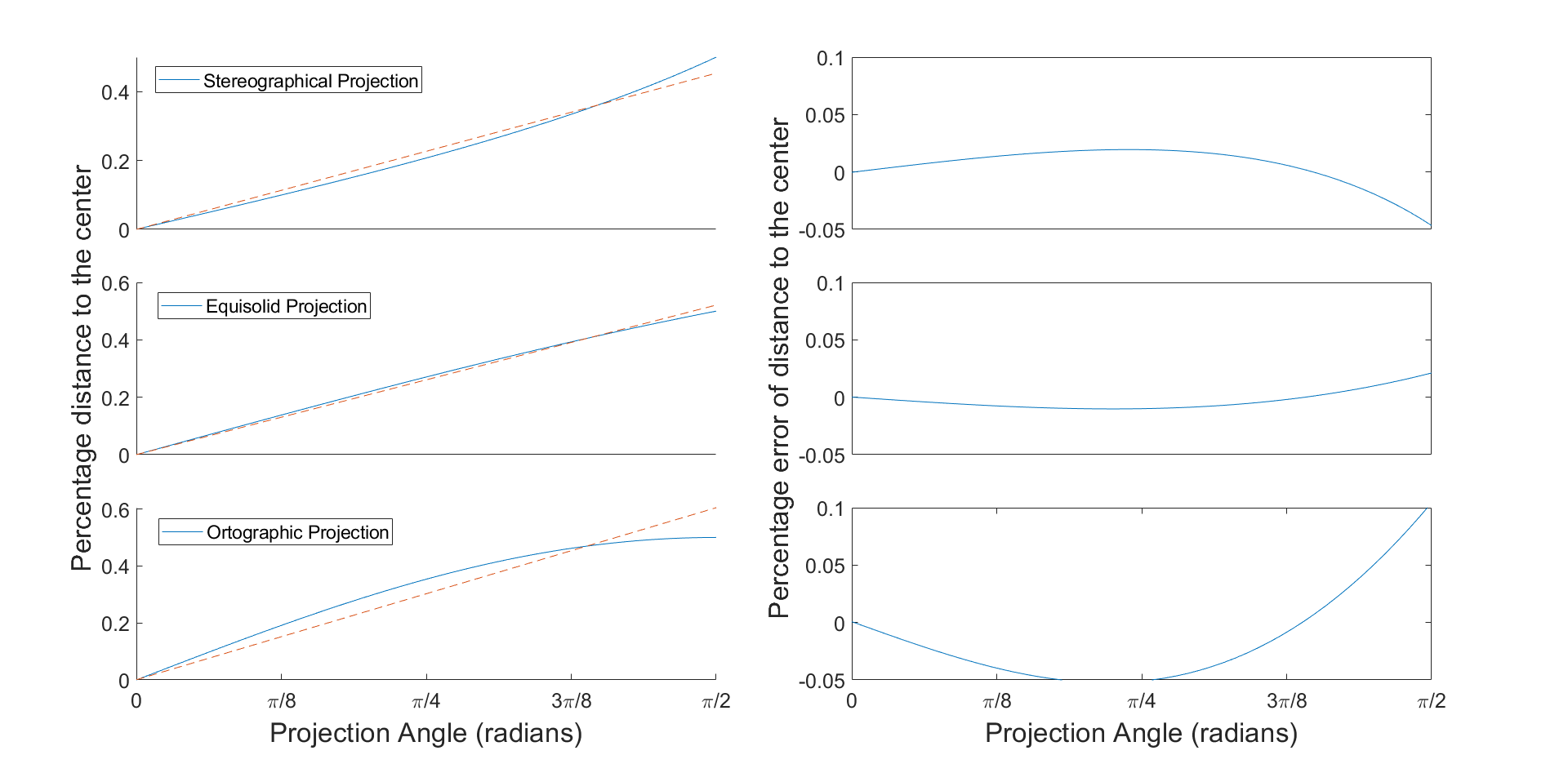}
    \caption{Different projections and respective linear approximations (left) and the percentage error (right).}
    \label{fig:Projecoes}
\end{figure}

Still, it might just be the case that the OPD camera has also an equidistant projection. We tried but were unable to find examples of all-sky cameras that provided their camera specifications of projections other than the equidistant. We could also not obtain a different all-sky camera for ourselves, and test on-site. So to confirm our assumptions, we checked how well these projections could be approximated by a linear function (see figure \ref{fig:Projecoes}). For all the listed projections, we can immediately notice that the error increases rapidly for higher projection angles (in the stereographical projection the error may surpass 2\% for angles higher than 80$^\circ$).

Although for the stereographical and equisolid case  the error is below the 2\% mark, with the ortographical case the error is higher. Fortunately, this last projection seems to be much less common as a default camera parameter. If the operator still wishes to consider the ortographical projection it can choose to overfit to the lower projection angles up a defined analysis threshold. Another option is to estimate iteratively  the projection values using the ZenithPhaseCalculator.py coupled with the graphical tool, but this is not an automated procedure.

\subsection{Cloud Mask}

\begin{figure}[!ht]
\centering
\begin{minipage}{.48\textwidth}
  \centering
  \includegraphics[width=70mm]{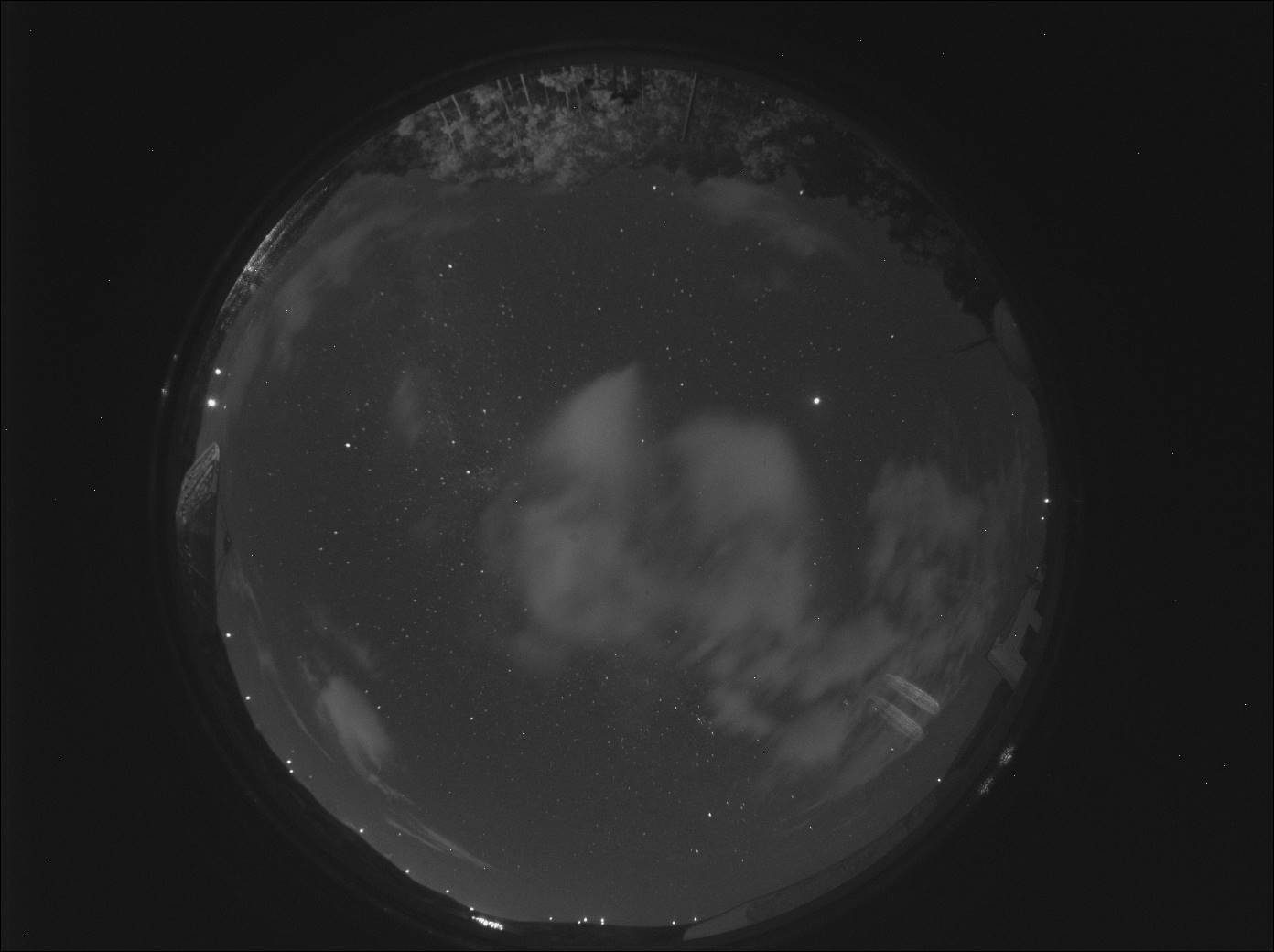}
\end{minipage}%
\begin{minipage}{.48\textwidth}
  \centering
  \setlength{\fboxsep}{0pt}
  \fbox{\includegraphics[width=70mm]{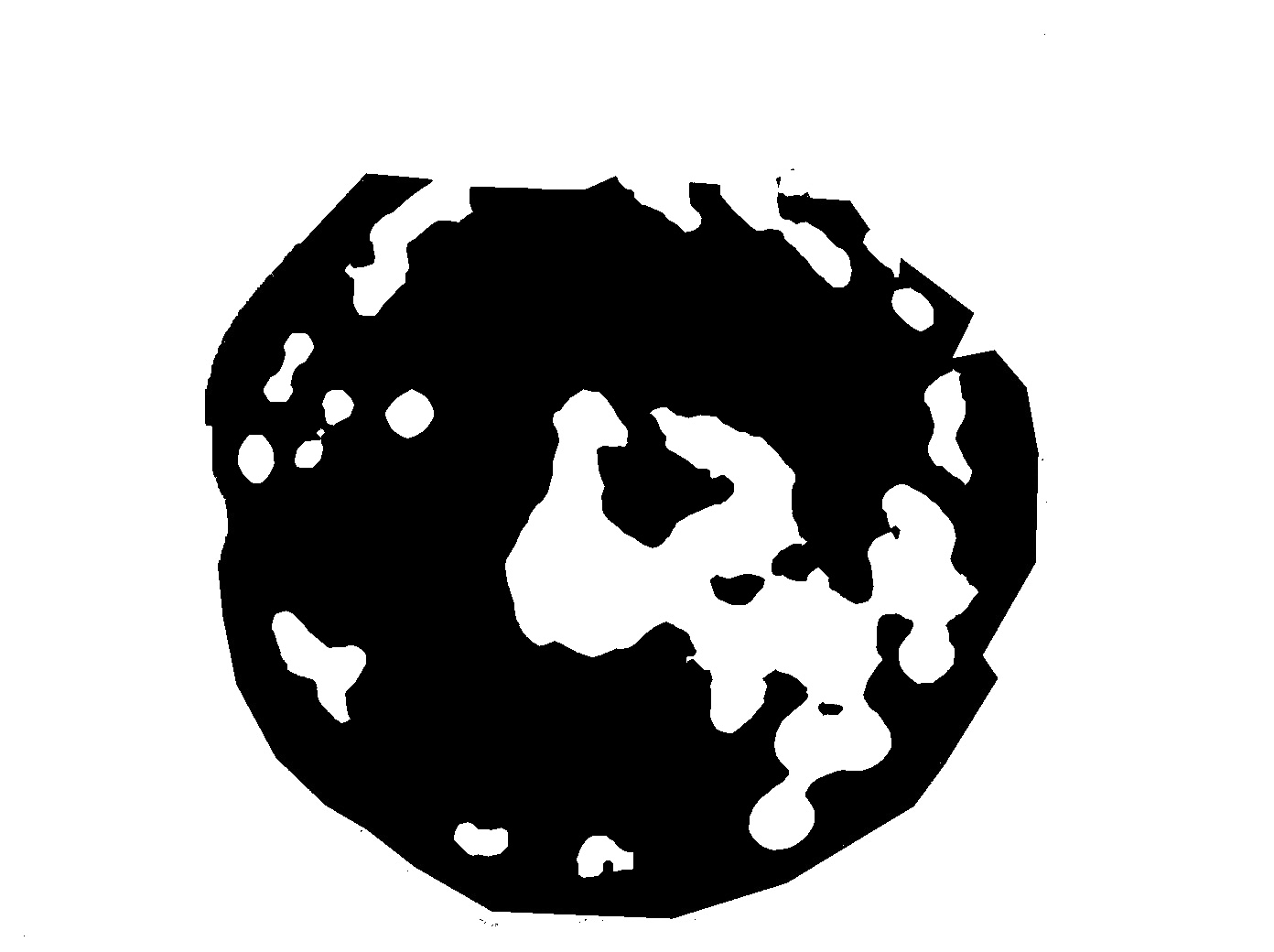}}
\end{minipage}
\caption{Visible cloud detected with our all-sky image from PASO (left) and corresponding cloud mask obtained with {\tt{CLOWN}} (right).}
\label{fig:visiblecloud}
\end{figure}

\begin{figure}[!ht]
\centering
\begin{minipage}{.48\textwidth}
  \centering
  \includegraphics[width=70mm]{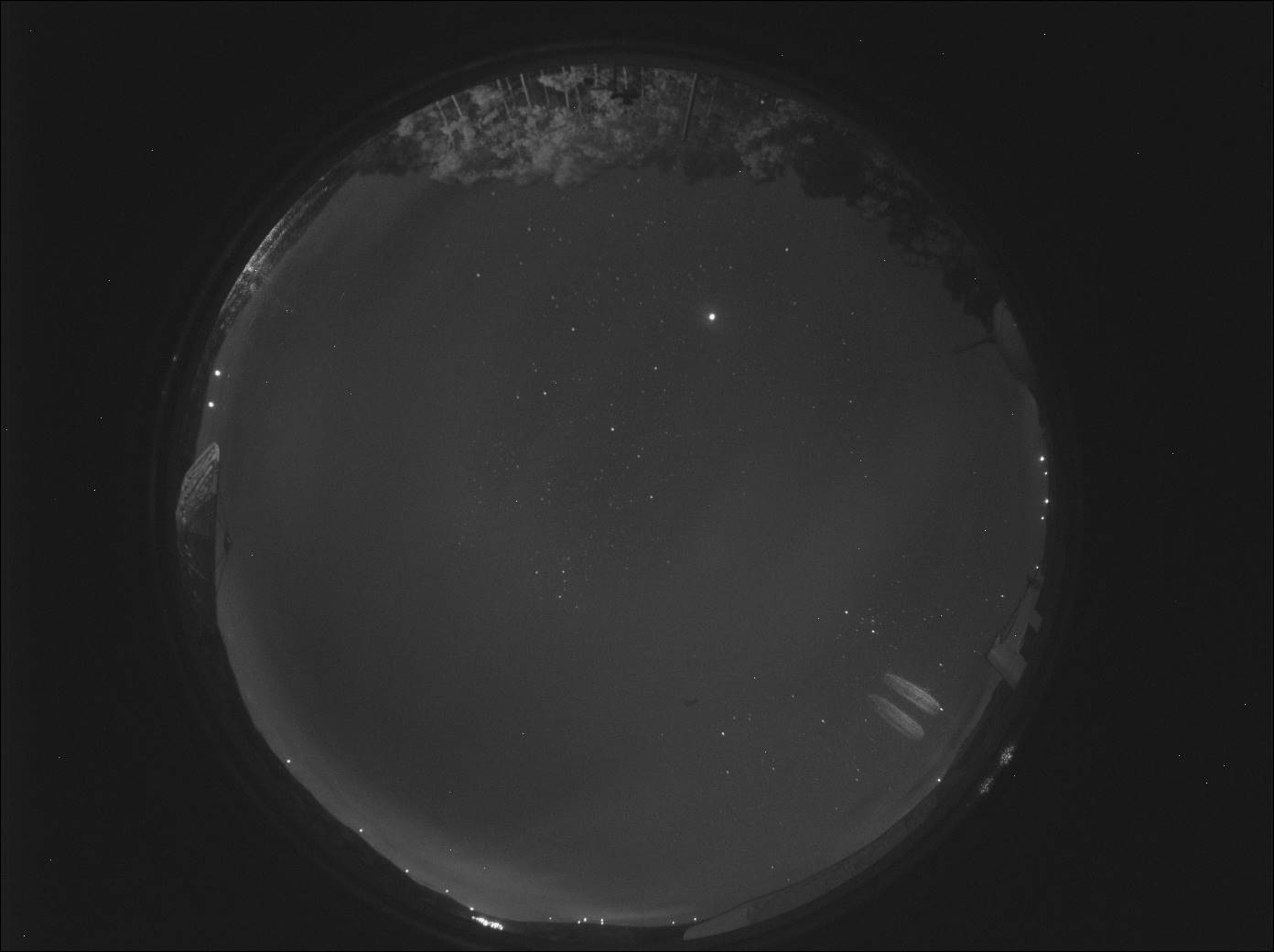}
\end{minipage}%
\begin{minipage}{.48\textwidth}
  \centering
  \setlength{\fboxsep}{0pt}
  \fbox{\includegraphics[width=70mm]{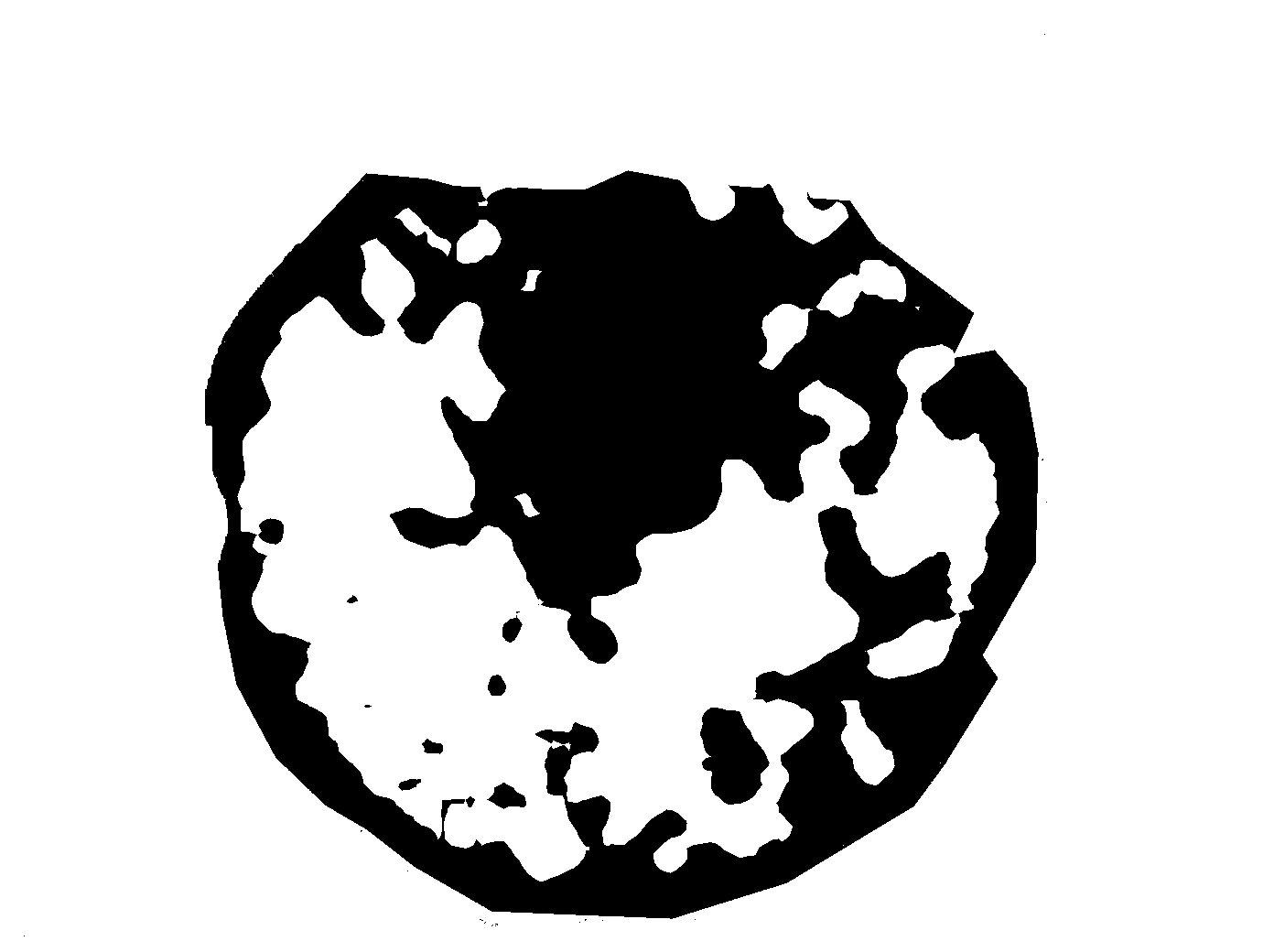}}
\end{minipage}
\caption{Invisible cloud detected with our all-sky image from PASO (left) and corresponding cloud mask obtained with {\tt{CLOWN}} (right). The telescope would still operate on these conditions, as it does not reach 75\% blocked sky.}
\label{fig:invisiblecloud}
\end{figure}

\begin{figure}[!ht]
\centering
\begin{minipage}{.48\textwidth}
  \centering
  \includegraphics[width=70mm]{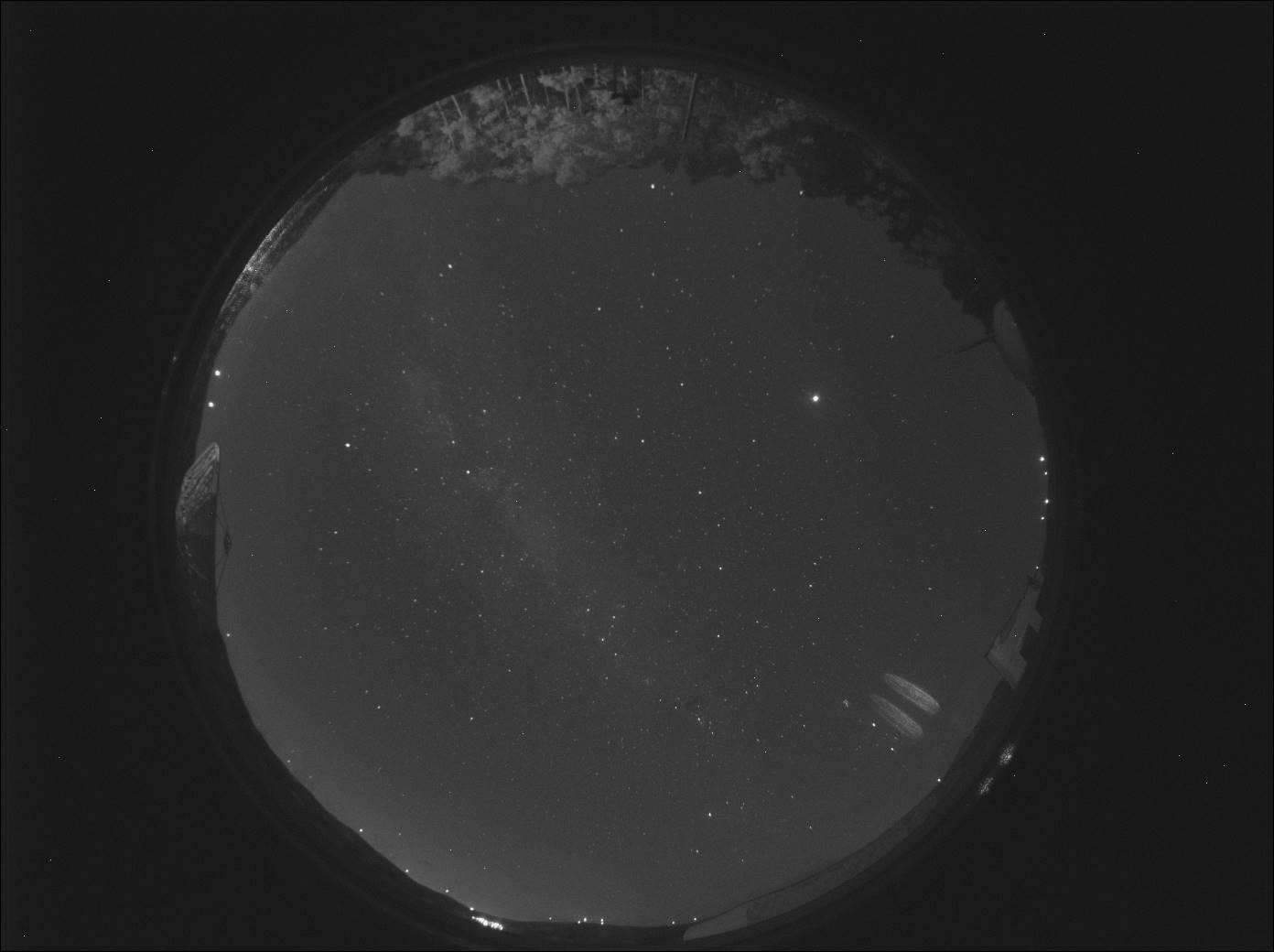}
\end{minipage}%
\begin{minipage}{.48\textwidth}
  \centering
  \setlength{\fboxsep}{0pt}
  \fbox{\includegraphics[width=70mm]{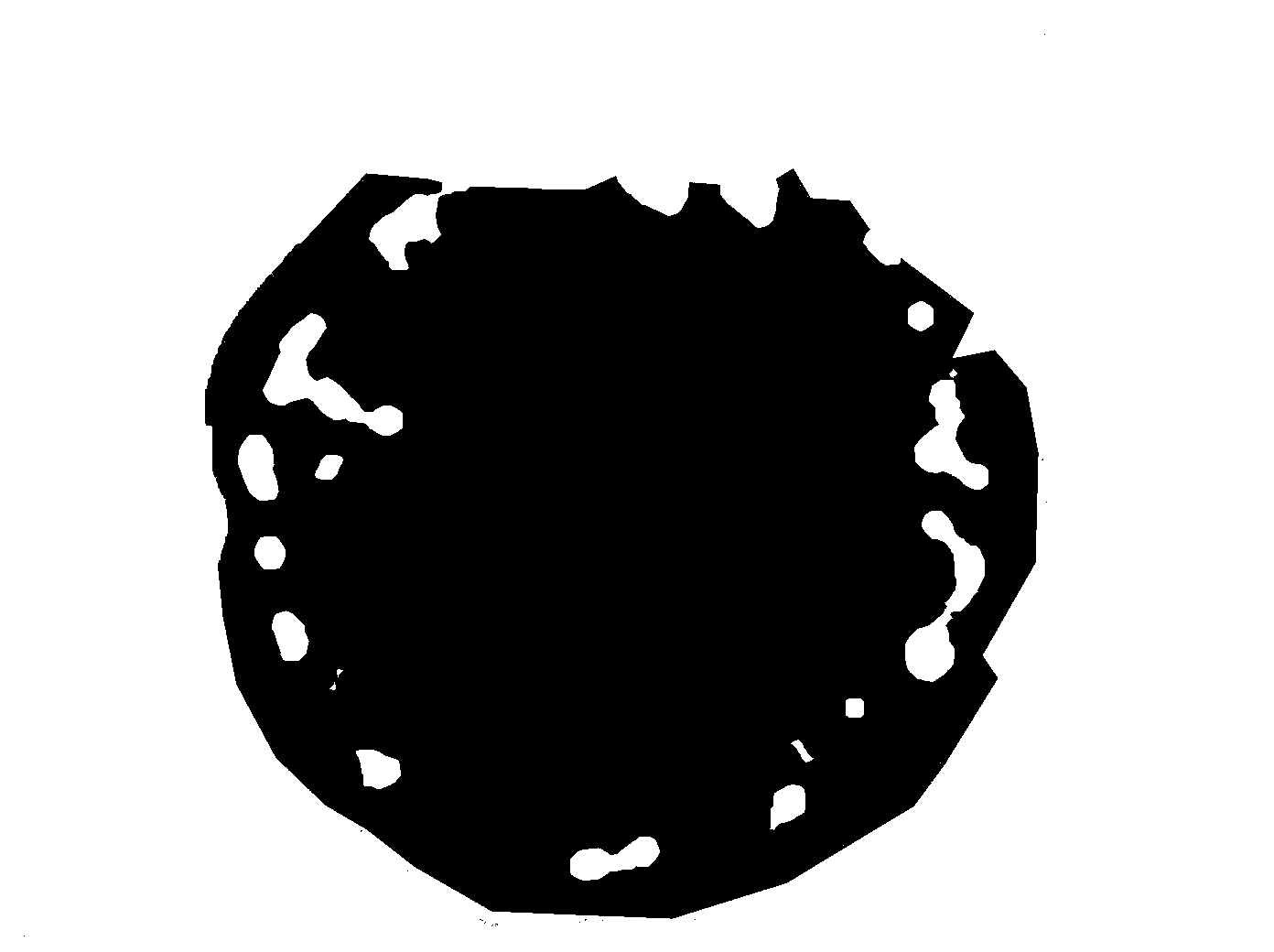}}
\end{minipage}
\caption{Clear sky image obtained with our all-sky image from PASO (left) and corresponding cloud mask obtained with {\tt{CLOWN}} (right).}
\label{fig:nocloud}
\end{figure}

\begin{figure}[!ht]
\centering
\begin{minipage}{.48\textwidth}
  \centering
  \includegraphics[width=70mm]{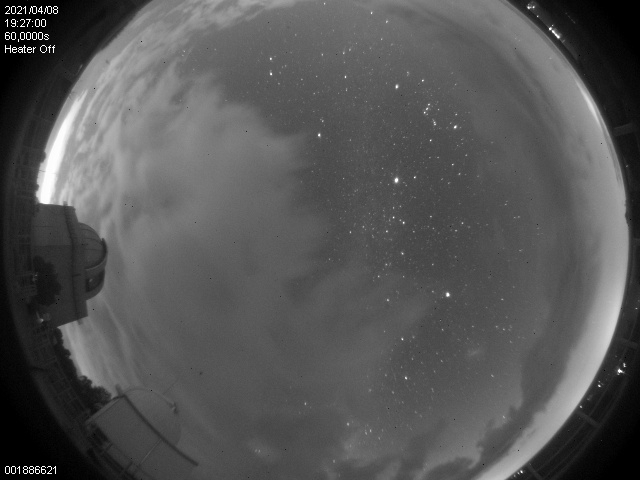}
\end{minipage}%
\begin{minipage}{.48\textwidth}
  \centering
  \setlength{\fboxsep}{0pt}
  \fbox{\includegraphics[width=70mm]{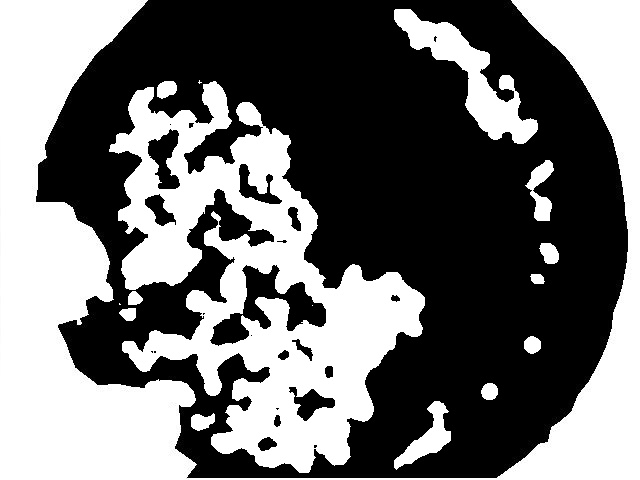}}
\end{minipage}
\caption{All-sky image taken from the OPD with cloudy conditions \cite{LNA} (left) and corresponding cloud mask obtained with {\tt{CLOWN}} (right).}
\label{fig:cloudLNA}
\end{figure}

The resulting cloud mask is saved as a binary image to remove ambiguities in the creation of an user threshold for what would be designed as an ``acceptable cloudiness" level. To produce a cloud mask a white blob is superposed on every missing star in the image. The blobs size was calculated by dividing the useful area of the image by the number of objects detected in the image. The overlap of these star blobs produces the final shape of the clouds. Although it was implemented to be an automated procedure, the size of the blobs can still be a free parameter controlled manually after an operator decision. Although we developed {\tt{CLOWN}} to be an automated procedure for our case study, it can be operated in a more manual control. In this case, the operator must be careful to avoid wrong cloud detections due to false positives. To correct for false positives and disable the software to wrongly block areas of clear sky, a local binarization procedure is also applied. For each pixel, the sum of the neighborhood pixels was compared with a threshold value. In our assumption, a cloud would block a large number of stars in the same area, and so if a star is the only one in that area, it may be considered a false-positive and removed. We tested this procedure by checking the values for the neighborhood size (in number of pixels) and the threshold value can be defined by the user in the configuration file. In figure \ref{fig:visiblecloud}, figure \ref{fig:invisiblecloud}, and in figure \ref{fig:nocloud} one can see examples of cloud masks obtained in different cloud conditions, as observed by the all-sky camera at PASO, and on figure \ref{fig:cloudLNA} we can see an example using the OPD all-sky camera.

\begin{figure}[!ht]
\centering
\begin{minipage}{.48\textwidth}
  \centering
  \includegraphics[width=70mm]{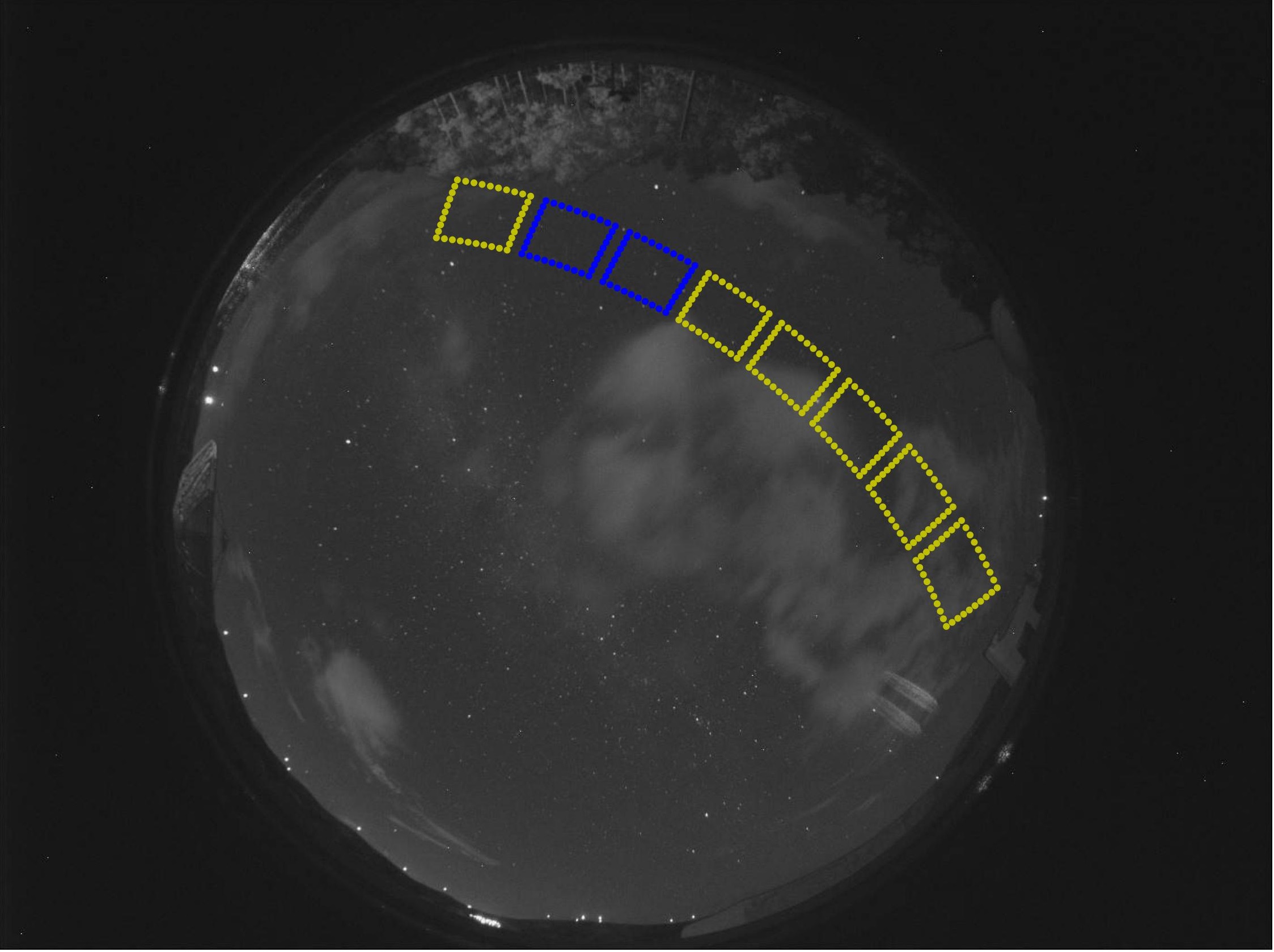}
\end{minipage}%
\begin{minipage}{.48\textwidth}
  \centering
  \includegraphics[width=70mm]{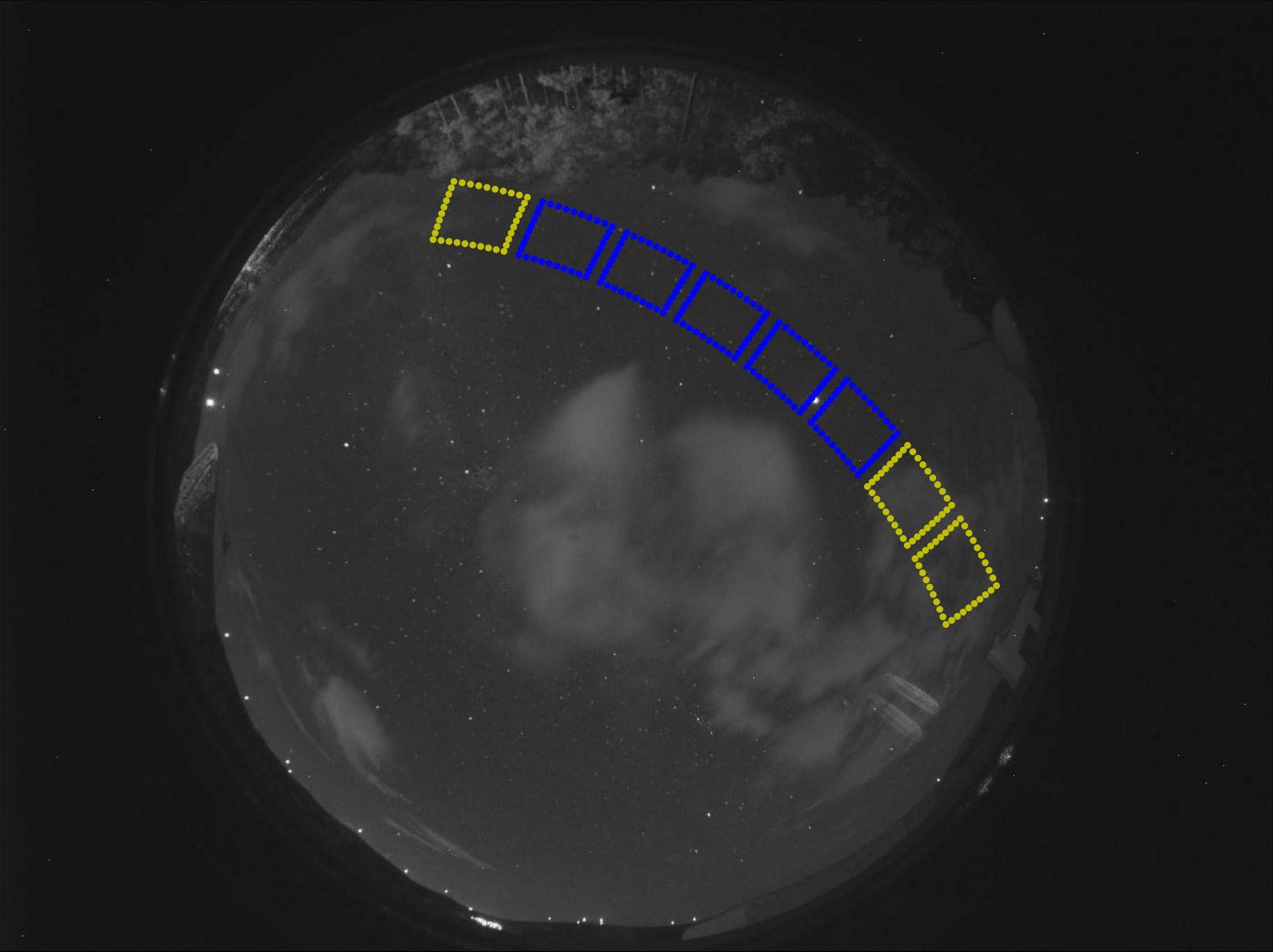}
\end{minipage}
\caption{In these images we can see a theoretical planned one night survey in the optical telescope where the rectangles indicate the observing areas of 6.9$^\circ\times6.9^\circ$ corresponding to the telescope FoV of 2.3$^\circ\times2.3^\circ$ plus 1 FoV of margin. Yellow and dark blue boundaries reveal areas with and without clouds, respectively. This information is then passed onto the telescope enabling it point to a cloudless area. Right and left images show the change in cloud conditions within a five minute interval.}
\label{fig:survey}
\end{figure}

\subsection{Running {\tt{CLOWN}} as a service }

The optical telescopes will execute every day a previously planned mission for the entire night (see example in figure \ref{fig:survey}). As seen in the example image, if the telescope operation starts from the image left side, it will acquire corrupted data due to cloud blocking, wasting survey time and causing processing hurdles. However, {\tt{CLOWN}} immediately detects the presence of clouds in that region and skips until it finds a clear-sky region. This operation increases the survey efficiency of the telescope in particular for space debris surveillance operations that require the acquisition of hundreds of images all night and thus avoid observation time lost as much as possible. {\tt{CLOWN}} enables a quick reaction and if cloud conditions change later, the telescopes can return to the area that was skipped beforehand and recover sky coverage as seen in figure \ref{fig:survey} (right). 

\begin{figure}[!ht]
    \centering
    \includegraphics[width=\textwidth]{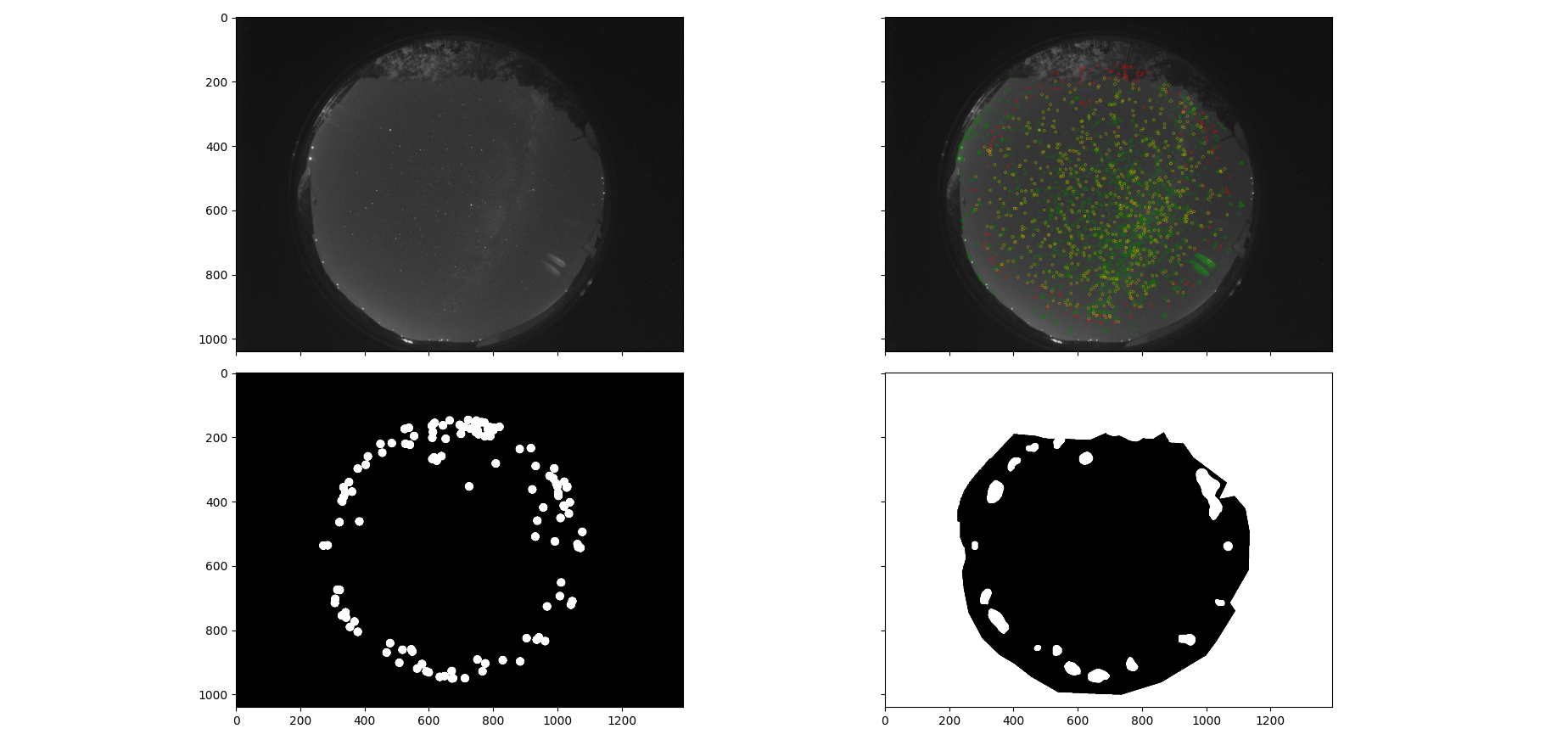}
    \caption{Graphical representation of image (top left), detected objects, expected stars and matched stars (top right), output of first analysis step (bottom left) and final output (bottom right).}
    \label{fig:representacao1}
\end{figure}

\begin{figure}[!ht]
    \centering
    \includegraphics[width=\textwidth]{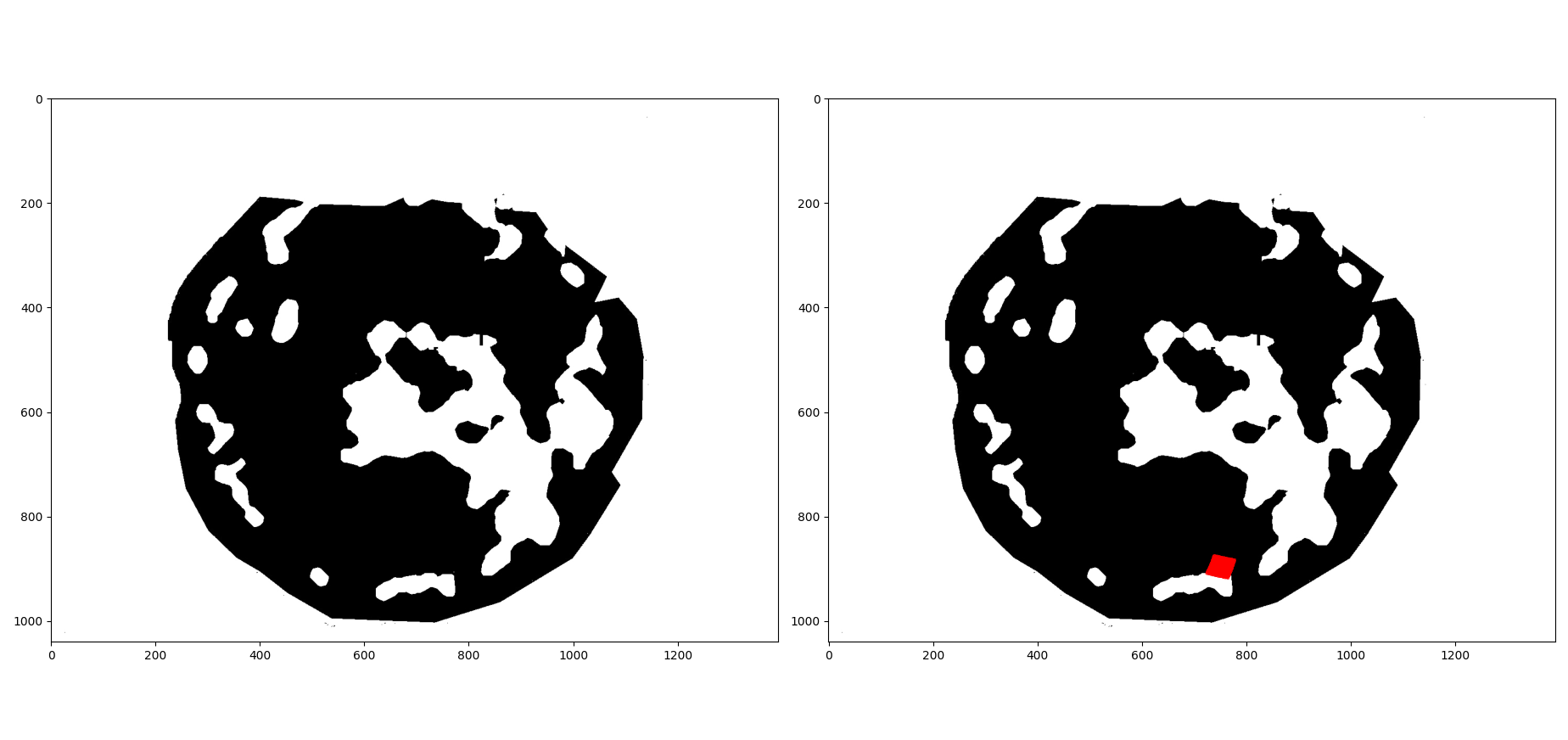}
    \caption{CheckPoint.py illustration: showing the area being checked.}
    \label{fig:representacao2}
\end{figure}

\begin{figure}[!ht]
    \centering
    \includegraphics[width=\textwidth]{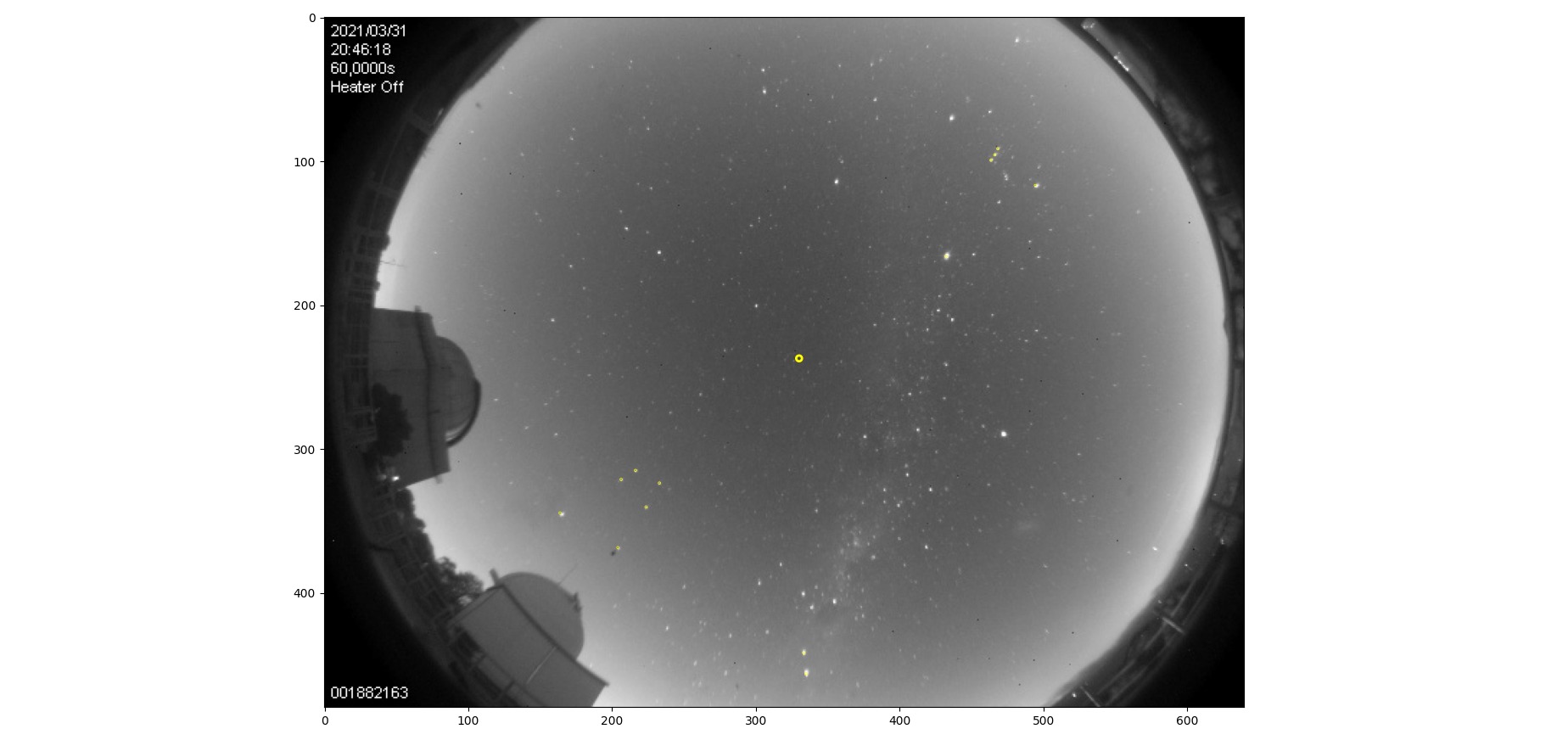}
    \caption{ZenithPointCalculator.py illustration. The zenith point is drawn in the center and the coordinates given in the AltAz file are drawn, so the user can visually see if the correspondence is correct or what error might be causing the mismatch.}
    \label{fig:representacao3}
\end{figure}

This program can also be used to guide user manual operation if required as it can correlate Ra/Dec coordinates to pixels in the image providing an easier visualization of the clear night sky for the operator.  After {\tt{CLOWN}} installation\footnote{Code freely available on \href{https://github.com/LuisGoncalves1/CLOWN}{GitHub}} we can check the two built-in examples with OPD and PASO files to enable program testing and careful integration within an observatory control environment. The three {\tt{CLOWN}} executables (\textbf{CLOWN.py}, \textbf{CheckPoint.py} and \textbf{ZenithPhaseCalculator.py}) run in the {\tt{CLOWN}} directory according to the flowchart described in figure \ref{fig:flowchart} and do follow a Python command line approach:

\begin{description}

    \item[CLOWN.py] {corresponds to the core program, creating a cloud mask from an image. This tool assumes that all-sky images are correctly stored in the specified Images folder path indicated in the configuration file. This enables {\tt{CLOWN}} to be used on a larger system controlling multiple all-sky cameras with different configuration files on different locations remotely at the same time. If the GRAPH option is enabled in the configuration file, a visual representation is done for every image with the star distribution (found and not found stars) as shown in figure \ref{fig:representacao1}. This tool follows a Python command line approach or a dashboard widget:\\
\makebox[8cm][l]{\tt{python3  CLOWN.py  config\_file}} \\[5pt]

}

    \item[CheckPoint.py] {It checks the existence of clouds blocking for either one (Ra,Dec) coordinate pair or a list of (Ra,Dec) coordinate pairs in a specific cloud mask and can be run as shown below respectively. In the CheckPoint tool, there is also a visual representation for easy understanding as shown in figure \ref{fig:representacao2}. This can be disabled in the GRAPH parameter in the configuration file.}

\makebox[8cm][l]{\tt{python3  CheckPoint.py  config\_file  MaskPath ~ Ra ~ Dec}} \\[5pt]
\makebox[8cm][l]{\tt{python3  CheckPoint.py  config\_file  MaskPath RaDecFile}} \\[5pt]

\item[ZenithPhaseCalculator.py]{The ZenithPhaseCalculator.py calculates the phase and zenith corrections as explained in section \ref{sec:StarMatching} and can be run as shown below. The AltAzFile and XYFile contain the corresponding coordinates for several stars in the image field, which can also be obtained with a general-purpose astronomy software like  Stellarium. The Area parameter controls the area around the center to search for the zenith. This tool also has visual assistance as shown in figure \ref{fig:representacao3}.

\makebox[8cm][l]{\tt{python3  ZenithPhaseCalculator.py AltAzFile XYFile Area}}} \\[5pt]

\end{description}

\subsection{Results}

{\tt{CLOWN}} was extensively tested with many acquired images in different cloud conditions. The software showed its pipeline can detect the presence of clouds with precision and provide a cloud monitoring service to astronomical operation. We presented examples of the results obtained for various types of cloud conditions (visible clouds in figure \ref{fig:visiblecloud}, invisible clouds in figure \ref{fig:invisiblecloud}, and no clouds in figure \ref{fig:nocloud}). {\tt{CLOWN}} was also able to analyze images without known camera model specifications as can be seen in figure \ref{fig:cloudLNA} - an image obtained by the all-sky camera located at OPD, Brazil. The existence of ``holes" in the mask is not an issue as they are smaller than the telescope FoV (2.3$^\circ\times2.3^\circ$) we are considering. In every tested mask, there were several “clouds” near the horizon (below 30 degrees of altitude), which for the application purposes was not considered a problem, since near horizon areas are not of interest to our main observation cases. For images with more than 75\% of the sky blocked by clouds, the telescope would automatically shut the dome to protect the equipment from potential damage. {\tt{CLOWN}} is also very effective to produce cloud maps of very bright clouds illuminated by the Moon, which would normally cause a large number of false detections.

\section{Conclusion}

{\tt{CLOWN}} is a new open-source tool to optimize large sky area survey observations meant specially for SSA monitoring and in our case for surveys targeting LEO mega-constellations and space-debris surveys. CLOWN was developed as an observatory tool coupled to the PASO optical telescopes. The code is freely available at \href{https://github.com/LuisGoncalves1/CLOWN}{GitHub}. {\tt{CLOWN}} can also be used as a ``Cloud Watch as a Service" to increase the efficiency of active survey missions other than space debris: it can monitor, in real-time and automatically, the position of clouds in the entire sky which enables the automatic use of an optical telescope even in cloudy conditions.

The program was developed with a low computational cost and applies to any type of all-sky camera even when there is no a priori knowledge of camera specifications, as was tested with the Brazil OPD images. For a well known camera with model specifications available extensive testing shows that {\tt{CLOWN}} obtained star coordinates with a median error of 0.42 $^\circ$; however, if there is not a priori knowledge of camera model specifications the median error is 0.61$^\circ$. Even for cases where the moon illuminates the clouds and negatively affects our images, CLOWN
is still robust enough to create a correct cloud mask.

Our all-sky camera is already installed in PASO and is accessed via a visual assistance dashboard tool enabling automatic and also manual operations.

For future improvements, we plan to explore machine learning tools for cloud detection when a sufficiently high amount of images for algorithm training becomes available.  This may produce faster and even better results. The images could also be used for statistical studies on local night cloudy conditions and improve the local monitoring on the number of cloudy nights, an important parameter for certification of Dark Sky astrotourism destinations. Indeed, PASO is located at Aldeias do Xisto Dark Sky Tourism destination certified by the by the Starlight Foundation and the data collected will be of importance for certification procedures and dark sky quality control.

Images of our all-sky camera will be made available online in the near future, updated every minute, following the example of \cite{LNA}.


\begin{acknowledgments}
    The team acknowledges support from ENGAGE-SKA Research Infrastructure, ref. POCI-01-0145-FEDER-022217 funded by COMPETE 2020 and FCT – Fundação para a Ciência e a Tecnologia, I.P., Portugal; Project Lab. Associado UID/EEA/50008/2019; Project Centro de Investigação em Ciências Geo-Espaciais, reference UIDB/00190/2020, funded by COMPETE 2020 and FCT, Portugal; project UIDB/00194/2020 ({\tt https://doi.org/10.54499/UIDB/00194/2020}) and UIDP/00194/2020 ({\tt https://doi.org/10.54499/UIDP/00194/2020}) funded by FCT, IP., Portugal. The team acknowledges support by the European Space Surveillance \& Tracking network, under the grant agreement 2-3SST2018-20 funded by European Commission H2020 Program. The team  also thanks José Silva for improving PASO's all-sky camera assembly. We would like to thank the suggestions and comments from the referee that helped to much improve our manuscript.
\end{acknowledgments}

\textbf{Used Softwares} : Astropy, AllSkEye, scikit-image, Stellarium



\bibliography{sample631}{}

\begin{thebibliography}{}
\expandafter\ifx\csname natexlab\endcsname\relax\def\natexlab#1{#1}\fi
\providecommand{\url}[1]{\href{#1}{#1}}
\providecommand{\dodoi}[1]{doi:~\href{http://doi.org/#1}{\nolinkurl{#1}}}
\providecommand{\doeprint}[1]{\href{http://ascl.net/#1}{\nolinkurl{http://ascl.net/#1}}}
\providecommand{\doarXiv}[1]{\href{https://arxiv.org/abs/#1}{\nolinkurl{https://arxiv.org/abs/#1}}}

\bibitem[{LNA(2022)}]{LNA}
 2022, Observatório do Pico dos Dias.
\newblock \url{http://200.131.64.207/allsky/centralclima.html}

\bibitem[{Adimurthy \& Ganeshan(2006)}]{ADIMURTHY2006168}
Adimurthy, V., \& Ganeshan, A. 2006, Acta Astronautica, 58, 168,
  \dodoi{https://doi.org/10.1016/j.actaastro.2005.09.002}

\bibitem[{{Astropy Collaboration} {et~al.}(2013){Astropy Collaboration},
  {Robitaille}, {Tollerud}, {Greenfield}, {Droettboom}, {Bray}, {Aldcroft},
  {Davis}, {Ginsburg}, {Price-Whelan}, {Kerzendorf}, {Conley}, {Crighton},
  {Barbary}, {Muna}, {Ferguson}, {Grollier}, {Parikh}, {Nair}, {Unther},
  {Deil}, {Woillez}, {Conseil}, {Kramer}, {Turner}, {Singer}, {Fox}, {Weaver},
  {Zabalza}, {Edwards}, {Azalee Bostroem}, {Burke}, {Casey}, {Crawford},
  {Dencheva}, {Ely}, {Jenness}, {Labrie}, {Lim}, {Pierfederici}, {Pontzen},
  {Ptak}, {Refsdal}, {Servillat}, \& {Streicher}}]{astropy:2013}
{Astropy Collaboration}, {Robitaille}, T.~P., {Tollerud}, E.~J., {et~al.} 2013,
  \aap, 558, A33, \dodoi{10.1051/0004-6361/201322068}

\bibitem[{{Astropy Collaboration} {et~al.}(2018){Astropy Collaboration},
  {Price-Whelan}, {Sip{\H{o}}cz}, {G{\"u}nther}, {Lim}, {Crawford}, {Conseil},
  {Shupe}, {Craig}, {Dencheva}, {Ginsburg}, {Vand erPlas}, {Bradley},
  {P{\'e}rez-Su{\'a}rez}, {de Val-Borro}, {Aldcroft}, {Cruz}, {Robitaille},
  {Tollerud}, {Ardelean}, {Babej}, {Bach}, {Bachetti}, {Bakanov}, {Bamford},
  {Barentsen}, {Barmby}, {Baumbach}, {Berry}, {Biscani}, {Boquien}, {Bostroem},
  {Bouma}, {Brammer}, {Bray}, {Breytenbach}, {Buddelmeijer}, {Burke},
  {Calderone}, {Cano Rodr{\'\i}guez}, {Cara}, {Cardoso}, {Cheedella}, {Copin},
  {Corrales}, {Crichton}, {D'Avella}, {Deil}, {Depagne}, {Dietrich}, {Donath},
  {Droettboom}, {Earl}, {Erben}, {Fabbro}, {Ferreira}, {Finethy}, {Fox},
  {Garrison}, {Gibbons}, {Goldstein}, {Gommers}, {Greco}, {Greenfield},
  {Groener}, {Grollier}, {Hagen}, {Hirst}, {Homeier}, {Horton}, {Hosseinzadeh},
  {Hu}, {Hunkeler}, {Ivezi{\'c}}, {Jain}, {Jenness}, {Kanarek}, {Kendrew},
  {Kern}, {Kerzendorf}, {Khvalko}, {King}, {Kirkby}, {Kulkarni}, {Kumar},
  {Lee}, {Lenz}, {Littlefair}, {Ma}, {Macleod}, {Mastropietro}, {McCully},
  {Montagnac}, {Morris}, {Mueller}, {Mumford}, {Muna}, {Murphy}, {Nelson},
  {Nguyen}, {Ninan}, {N{\"o}the}, {Ogaz}, {Oh}, {Parejko}, {Parley}, {Pascual},
  {Patil}, {Patil}, {Plunkett}, {Prochaska}, {Rastogi}, {Reddy Janga},
  {Sabater}, {Sakurikar}, {Seifert}, {Sherbert}, {Sherwood-Taylor}, {Shih},
  {Sick}, {Silbiger}, {Singanamalla}, {Singer}, {Sladen}, {Sooley},
  {Sornarajah}, {Streicher}, {Teuben}, {Thomas}, {Tremblay}, {Turner},
  {Terr{\'o}n}, {van Kerkwijk}, {de la Vega}, {Watkins}, {Weaver}, {Whitmore},
  {Woillez}, {Zabalza}, \& {Astropy Contributors}}]{astropy:2018}
{Astropy Collaboration}, {Price-Whelan}, A.~M., {Sip{\H{o}}cz}, B.~M., {et~al.}
  2018, \aj, 156, 123, \dodoi{10.3847/1538-3881/aabc4f}

\bibitem[{{Aydi} {et~al.}(2022){Aydi}, {Sokolovsky}, {Bright}, {Tremou},
  {Nyamai}, {Evans}, {Strader}, {Chomiuk}, {Myers}, {Hambsch}, {Page},
  {Buckley}, {Woodward}, {Walter}, {Mr{\'o}z}, {Vallely}, {Geballe},
  {Banerjee}, {Gehrz}, {Fender}, {Gromadzki}, {Kawash}, {Knigge}, {Mukai},
  {Munari}, {Orio}, {Ribeiro}, {Sokoloski}, {Starrfield}, {Udalski}, \&
  {Woudt}}]{2022ApJ...939....6A}
{Aydi}, E., {Sokolovsky}, K.~V., {Bright}, J.~S., {et~al.} 2022, \apj, 939, 6,
  \dodoi{10.3847/1538-4357/ac913b}

\bibitem[{{Barentine} \& {Heim}(2023)}]{2023arXiv231102184B}
{Barentine}, J.~C., \& {Heim}, J. 2023, arXiv e-prints, arXiv:2311.02184,
  \dodoi{10.48550/arXiv.2311.02184}

\bibitem[{Bassett {et~al.}(2021)Bassett, Rapetti, Tauscher, Nhan, Bordenave,
  Hibbard, \& Burns}]{Bassett_2021}
Bassett, N., Rapetti, D., Tauscher, K., {et~al.} 2021, The Astrophysical
  Journal, 923, 33, \dodoi{10.3847/1538-4357/ac1cde}

\bibitem[{Beghdadi \& {Le Negrate}(1989)}]{BEGHDADI1989162}
Beghdadi, A., \& {Le Negrate}, A. 1989, Computer Vision, Graphics, and Image
  Processing, 46, 162, \dodoi{https://doi.org/10.1016/0734-189X(89)90166-7}

\bibitem[{{Ben{\'\i}tez} {et~al.}(2015){Ben{\'\i}tez}, {Dupke}, {Moles},
  {Sodr{\'e}}, {Cenarro}, {Mar{\'\i}n Franch}, {Taylor}, {Crist{\'o}bal},
  {Fern{\'a}ndez-Soto}, {Mendes de Oliveira}, {Cepa-Nogu{\'e}}, {Abramo},
  {Alcaniz}, {Overzier}, {Hern{\'a}ndez-Monteagudo}, {Alfaro}, {Kanaan},
  {Carvano}, {Reis}, \& {J-PAS Collaboration}}]{JPAS}
{Ben{\'\i}tez}, N., {Dupke}, R., {Moles}, M., {et~al.} 2015, in Highlights of
  Spanish Astrophysics VIII, 148--153

\bibitem[{Blum {et~al.}(2022)Blum, Digel, Drlica-Wagner, Habib, Heitmann,
  Ishak, Jha, Kahn, Mandelbaum, Marshall, Newman, Roodman, \& Stubbs}]{LSST}
Blum, B., Digel, S., Drlica-Wagner, A., {et~al.} 2022, Snowmass2021 Cosmic
  Frontier White Paper: Rubin Observatory after LSST.
\newblock \doarXiv{2203.07220}

\bibitem[{Bulmann {et~al.}(1974)Bulmann, Balogh, Ashdown, Gabbard, Smith, \&
  Snapp}]{bulmann}
Bulmann, R., Balogh, J., Ashdown, R., {et~al.} 1974

\bibitem[{{Cenarro} {et~al.}(2019){Cenarro}, {Moles},
  {Crist{\'o}bal-Hornillos}, {Mar{\'\i}n-Franch}, {Ederoclite}, {Varela},
  {L{\'o}pez-Sanjuan}, {Hern{\'a}ndez-Monteagudo}, {Angulo}, {V{\'a}zquez
  Rami{\'o}}, {Viironen}, {Bonoli}, {Orsi}, {Hurier}, {San Roman}, {Greisel},
  {Vilella-Rojo}, {D{\'\i}az-Garc{\'\i}a}, {Logro{\~n}o-Garc{\'\i}a},
  {Gurung-L{\'o}pez}, {Spinoso}, {Izquierdo-Villalba}, {Aguerri}, {Allende
  Prieto}, {Bonatto}, {Carvano}, {Chies-Santos}, {Daflon}, {Dupke},
  {Falc{\'o}n-Barroso}, {Gon{\c{c}}alves}, {Jim{\'e}nez-Teja}, {Molino},
  {Placco}, {Solano}, {Whitten}, {Abril}, {Ant{\'o}n}, {Bello}, {Bielsa de
  Toledo}, {Castillo-Ram{\'\i}rez}, {Chueca}, {Civera},
  {D{\'\i}az-Mart{\'\i}n}, {Dom{\'\i}nguez-Mart{\'\i}nez},
  {Garzar{\'a}n-Calderaro}, {Hern{\'a}ndez-Fuertes}, {Iglesias-Marzoa},
  {I{\~n}iguez}, {Jim{\'e}nez Ruiz}, {Kruuse}, {Lamadrid}, {Lasso-Cabrera},
  {L{\'o}pez-Alegre}, {L{\'o}pez-Sainz}, {Ma{\'\i}cas}, {Moreno-Signes},
  {Muniesa}, {Rodr{\'\i}guez-Llano}, {Rueda-Teruel}, {Rueda-Teruel},
  {Soriano-Lagu{\'\i}a}, {Tilve}, {Valdivielso}, {Yanes-D{\'\i}az}, {Alcaniz},
  {Mendes de Oliveira}, {Sodr{\'e}}, {Coelho}, {Lopes de Oliveira}, {Tamm},
  {Xavier}, {Abramo}, {Akras}, {Alfaro}, {Alvarez-Candal}, {Ascaso}, {Beasley},
  {Beers}, {Borges Fernandes}, {Bruzual}, {Buzzo}, {Carrasco}, {Cepa},
  {Cortesi}, {Costa-Duarte}, {De Pr{\'a}}, {Favole}, {Galarza}, {Galbany},
  {Garcia}, {Gonz{\'a}lez Delgado}, {Gonz{\'a}lez-Serrano},
  {Guti{\'e}rrez-Soto}, {Hernandez-Jimenez}, {Kanaan}, {Kuncarayakti},
  {Landim}, {Laur}, {Licandro}, {Lima Neto}, {Lyman}, {Ma{\'\i}z
  Apell{\'a}niz}, {Miralda-Escud{\'e}}, {Morate}, {Nogueira-Cavalcante},
  {Novais}, {Oncins}, {Oteo}, {Overzier}, {Pereira}, {Rebassa-Mansergas},
  {Reis}, {Roig}, {Sako}, {Salvador-Rusi{\~n}ol}, {Sampedro},
  {S{\'a}nchez-Bl{\'a}zquez}, {Santos}, {Schmidtobreick}, {Siffert}, {Telles},
  \& {Vilchez}}]{JPlus}
{Cenarro}, A.~J., {Moles}, M., {Crist{\'o}bal-Hornillos}, D., {et~al.} 2019,
  \aap, 622, A176, \dodoi{10.1051/0004-6361/201833036}

\bibitem[{Chaple {et~al.}(2015)Chaple, Daruwala, \& Gofane}]{7095920}
Chaple, G.~N., Daruwala, R.~D., \& Gofane, M.~S. 2015, in 2015 International
  Conference on Technologies for Sustainable Development (ICTSD), 1--4,
  \dodoi{10.1109/ICTSD.2015.7095920}

\bibitem[{{Chomiuk} {et~al.}(2022){Chomiuk}, {Linford}, {Aydi}, {Bannister},
  {Krauss}, {Mioduszewski}, {Mukai}, {Nelson}, {Rupen}, {Ryder}, {Sokoloski},
  {Sokolovsky}, {Strader}, {Filipovic}, {Finzell}, {Kawash}, {Kool}, {Metzger},
  {Nyamai}, {Ribeiro}, {Roy}, {Urquhart}, \& {Weston}}]{2022yCat..22570049C}
{Chomiuk}, L., {Linford}, J.~D., {Aydi}, E., {et~al.} 2022, VizieR Online Data
  Catalog, J/ApJS/257/49, \dodoi{10.26093/cds/vizier.22570049}

\bibitem[{Coelho {et~al.}(2021)Coelho, Barbosa, Bergano, Correia, Freitas,
  Marques, Pandeirada, \& Ribeiro}]{coelho2021new}
Coelho, B., Barbosa, D., Bergano, M., {et~al.} 2021, in Proc. 8th European
  Conference on Space Debris (virtual), Darmstadt, Germany, 20–23 April 2021,
  ed. T.~Flohrer, S.~Lemmens, \& F.~Schmitz (ESA Space Debris Office).
\newblock
  \url{https://conference.sdo.esoc.esa.int/proceedings/sdc8/paper/290/SDC8-paper290.pdf}

\bibitem[{{Dark Energy Survey Collaboration} {et~al.}(2016){Dark Energy Survey
  Collaboration}, {Abbott}, {Abdalla}, {Aleksi{\'c}}, {Allam}, {Amara},
  {Bacon}, {Balbinot}, {Banerji}, {Bechtol}, {Benoit-L{\'e}vy}, {Bernstein},
  {Bertin}, {Blazek}, {Bonnett}, {Bridle}, {Brooks}, {Brunner}, {Buckley-Geer},
  {Burke}, {Caminha}, {Capozzi}, {Carlsen}, {Carnero-Rosell}, {Carollo},
  {Carrasco-Kind}, {Carretero}, {Castander}, {Clerkin}, {Collett}, {Conselice},
  {Crocce}, {Cunha}, {D'Andrea}, {da Costa}, {Davis}, {Desai}, {Diehl},
  {Dietrich}, {Dodelson}, {Doel}, {Drlica-Wagner}, {Estrada}, {Etherington},
  {Evrard}, {Fabbri}, {Finley}, {Flaugher}, {Foley}, {Fosalba}, {Frieman},
  {Garc{\'\i}a-Bellido}, {Gaztanaga}, {Gerdes}, {Giannantonio}, {Goldstein},
  {Gruen}, {Gruendl}, {Guarnieri}, {Gutierrez}, {Hartley}, {Honscheid}, {Jain},
  {James}, {Jeltema}, {Jouvel}, {Kessler}, {King}, {Kirk}, {Kron}, {Kuehn},
  {Kuropatkin}, {Lahav}, {Li}, {Lima}, {Lin}, {Maia}, {Makler}, {Manera},
  {Maraston}, {Marshall}, {Martini}, {McMahon}, {Melchior}, {Merson}, {Miller},
  {Miquel}, {Mohr}, {Morice-Atkinson}, {Naidoo}, {Neilsen}, {Nichol}, {Nord},
  {Ogando}, {Ostrovski}, {Palmese}, {Papadopoulos}, {Peiris}, {Peoples},
  {Percival}, {Plazas}, {Reed}, {Refregier}, {Romer}, {Roodman}, {Ross},
  {Rozo}, {Rykoff}, {Sadeh}, {Sako}, {S{\'a}nchez}, {Sanchez}, {Santiago},
  {Scarpine}, {Schubnell}, {Sevilla-Noarbe}, {Sheldon}, {Smith}, {Smith},
  {Soares-Santos}, {Sobreira}, {Soumagnac}, {Suchyta}, {Sullivan}, {Swanson},
  {Tarle}, {Thaler}, {Thomas}, {Thomas}, {Tucker}, {Vieira}, {Vikram},
  {Walker}, {Wechsler}, {Weller}, {Wester}, {Whiteway}, {Wilcox}, {Yanny},
  {Zhang}, \& {Zuntz}}]{DES}
{Dark Energy Survey Collaboration}, {Abbott}, T., {Abdalla}, F.~B., {et~al.}
  2016, \mnras, 460, 1270, \dodoi{10.1093/mnras/stw641}

\bibitem[{{Darnley} {et~al.}(2019){Darnley}, {Hounsell}, {O'Brien}, {Henze},
  {Rodr{\'\i}guez-Gil}, {Shafter}, {Shara}, {Vaytet}, {Bode}, {Ciardullo},
  {Davis}, {Galera-Rosillo}, {Harman}, {Harvey}, {Healy}, {Ness}, {Ribeiro}, \&
  {Williams}}]{2019Natur.565..460D}
{Darnley}, M.~J., {Hounsell}, R., {O'Brien}, T.~J., {et~al.} 2019, \nat, 565,
  460, \dodoi{10.1038/s41586-018-0825-4}

\bibitem[{Du(2017)}]{DU20178}
Du, R. 2017, Space Policy, 42, 8,
  \dodoi{https://doi.org/10.1016/j.spacepol.2017.10.005}

\bibitem[{Fleck(1995)}]{lentes}
Fleck, M.~M. 1995, IEEE Transactions on Reliability.
\newblock \url{https://api.semanticscholar.org/CorpusID:17158756}

\bibitem[{Froehlich {et~al.}(2020)Froehlich, Soria, \&
  De~Marchi}]{froehlich2020space}
Froehlich, A., Soria, D. A.~A., \& De~Marchi, E. 2020, Space Supporting Latin
  America: Latin America's Emerging Space Middle Powers, Vol.~25 (Springer
  Nature)

\bibitem[{Gonzalez \& Wood(2008)}]{Gonzalez}
Gonzalez, R., \& Wood, R. 2008, Digital Image Processing (Addison–Wesley
  Longman Publishing Co., Inc)

\bibitem[{Isar {et~al.}(2011)Isar, Firoiu, Nafornita, \& Moga}]{Isar11}
Isar, A., Firoiu, I., Nafornita, C., \& Moga, S. 2011, in Sonar Systems, ed.
  N.~Z. Kolev (Rijeka: IntechOpen), \dodoi{10.5772/19190}

\bibitem[{{Ivezi{\'c}} {et~al.}(2019){Ivezi{\'c}}, {Kahn}, {Tyson}, {Abel},
  {Acosta}, {Allsman}, {Alonso}, {AlSayyad}, {Anderson}, {Andrew}, {Angel},
  {Angeli}, {Ansari}, {Antilogus}, {Araujo}, {Armstrong}, {Arndt}, {Astier},
  {Aubourg}, {Auza}, {Axelrod}, {Bard}, {Barr}, {Barrau}, {Bartlett}, {Bauer},
  {Bauman}, {Baumont}, {Bechtol}, {Bechtol}, {Becker}, {Becla}, {Beldica},
  {Bellavia}, {Bianco}, {Biswas}, {Blanc}, {Blazek}, {Blandford}, {Bloom},
  {Bogart}, {Bond}, {Booth}, {Borgland}, {Borne}, {Bosch}, {Boutigny},
  {Brackett}, {Bradshaw}, {Brandt}, {Brown}, {Bullock}, {Burchat}, {Burke},
  {Cagnoli}, {Calabrese}, {Callahan}, {Callen}, {Carlin}, {Carlson},
  {Chandrasekharan}, {Charles-Emerson}, {Chesley}, {Cheu}, {Chiang}, {Chiang},
  {Chirino}, {Chow}, {Ciardi}, {Claver}, {Cohen-Tanugi}, {Cockrum}, {Coles},
  {Connolly}, {Cook}, {Cooray}, {Covey}, {Cribbs}, {Cui}, {Cutri}, {Daly},
  {Daniel}, {Daruich}, {Daubard}, {Daues}, {Dawson}, {Delgado}, {Dellapenna},
  {de Peyster}, {de Val-Borro}, {Digel}, {Doherty}, {Dubois},
  {Dubois-Felsmann}, {Durech}, {Economou}, {Eifler}, {Eracleous}, {Emmons},
  {Fausti Neto}, {Ferguson}, {Figueroa}, {Fisher-Levine}, {Focke}, {Foss},
  {Frank}, {Freemon}, {Gangler}, {Gawiser}, {Geary}, {Gee}, {Geha}, {Gessner},
  {Gibson}, {Gilmore}, {Glanzman}, {Glick}, {Goldina}, {Goldstein}, {Goodenow},
  {Graham}, {Gressler}, {Gris}, {Guy}, {Guyonnet}, {Haller}, {Harris},
  {Hascall}, {Haupt}, {Hernandez}, {Herrmann}, {Hileman}, {Hoblitt}, {Hodgson},
  {Hogan}, {Howard}, {Huang}, {Huffer}, {Ingraham}, {Innes}, {Jacoby}, {Jain},
  {Jammes}, {Jee}, {Jenness}, {Jernigan}, {Jevremovi{\'c}}, {Johns}, {Johnson},
  {Johnson}, {Jones}, {Juramy-Gilles}, {Juri{\'c}}, {Kalirai}, {Kallivayalil},
  {Kalmbach}, {Kantor}, {Karst}, {Kasliwal}, {Kelly}, {Kessler}, {Kinnison},
  {Kirkby}, {Knox}, {Kotov}, {Krabbendam}, {Krughoff}, {Kub{\'a}nek},
  {Kuczewski}, {Kulkarni}, {Ku}, {Kurita}, {Lage}, {Lambert}, {Lange},
  {Langton}, {Le Guillou}, {Levine}, {Liang}, {Lim}, {Lintott}, {Long},
  {Lopez}, {Lotz}, {Lupton}, {Lust}, {MacArthur}, {Mahabal}, {Mandelbaum},
  {Markiewicz}, {Marsh}, {Marshall}, {Marshall}, {May}, {McKercher}, {McQueen},
  {Meyers}, {Migliore}, {Miller}, {Mills}, {Miraval}, {Moeyens}, {Moolekamp},
  {Monet}, {Moniez}, {Monkewitz}, {Montgomery}, {Morrison}, {Mueller},
  {Muller}, {Mu{\~n}oz Arancibia}, {Neill}, {Newbry}, {Nief}, {Nomerotski},
  {Nordby}, {O'Connor}, {Oliver}, {Olivier}, {Olsen}, {O'Mullane}, {Ortiz},
  {Osier}, {Owen}, {Pain}, {Palecek}, {Parejko}, {Parsons}, {Pease},
  {Peterson}, {Peterson}, {Petravick}, {Libby Petrick}, {Petry},
  {Pierfederici}, {Pietrowicz}, {Pike}, {Pinto}, {Plante}, {Plate}, {Plutchak},
  {Price}, {Prouza}, {Radeka}, {Rajagopal}, {Rasmussen}, {Regnault}, {Reil},
  {Reiss}, {Reuter}, {Ridgway}, {Riot}, {Ritz}, {Robinson}, {Roby}, {Roodman},
  {Rosing}, {Roucelle}, {Rumore}, {Russo}, {Saha}, {Sassolas}, {Schalk},
  {Schellart}, {Schindler}, {Schmidt}, {Schneider}, {Schneider}, {Schoening},
  {Schumacher}, {Schwamb}, {Sebag}, {Selvy}, {Sembroski}, {Seppala}, {Serio},
  {Serrano}, {Shaw}, {Shipsey}, {Sick}, {Silvestri}, {Slater}, {Smith},
  {Smith}, {Sobhani}, {Soldahl}, {Storrie-Lombardi}, {Stover}, {Strauss},
  {Street}, {Stubbs}, {Sullivan}, {Sweeney}, {Swinbank}, {Szalay}, {Takacs},
  {Tether}, {Thaler}, {Thayer}, {Thomas}, {Thornton}, {Thukral}, {Tice},
  {Trilling}, {Turri}, {Van Berg}, {Vanden Berk}, {Vetter}, {Virieux},
  {Vucina}, {Wahl}, {Walkowicz}, {Walsh}, {Walter}, {Wang}, {Wang}, {Warner},
  {Wiecha}, {Willman}, {Winters}, {Wittman}, {Wolff}, {Wood-Vasey}, {Wu},
  {Xin}, {Yoachim}, \& {Zhan}}]{2019ApJ...873..111I}
{Ivezi{\'c}}, {\v{Z}}., {Kahn}, S.~M., {Tyson}, J.~A., {et~al.} 2019, \apj,
  873, 111, \dodoi{10.3847/1538-4357/ab042c}

\bibitem[{Jiang {et~al.}(2022)Jiang, Liu, Yang, Kang, Fan, \& Li}]{Jiang_2022}
Jiang, P., Liu, C., Yang, W., {et~al.} 2022, The Astrophysical Journal
  Supplement Series, 259, 4, \dodoi{10.3847/1538-4365/ac458d}

\bibitem[{Kessler \& Cour-Palais(1978)}]{Kessler1978CollisionFO}
Kessler, D.~J., \& Cour-Palais, B.~G. 1978, Journal of Geophysical Research,
  83, 2637

\bibitem[{Kingsbury(1999)}]{kingsbury1999image}
Kingsbury, N. 1999, Philosophical Transactions of the Royal Society of London.
  Series A: Mathematical, Physical and Engineering Sciences, 357, 2543

\bibitem[{Loza {et~al.}(2013)Loza, Bull, Hill, \& Achim}]{Loza}
Loza, A., Bull, D., Hill, P., \& Achim, A. 2013, Digital Signal Processing, 23,
  1856–1866, \dodoi{10.1016/j.dsp.2013.06.002}

\bibitem[{{Marr} \& {Hildreth}(1980)}]{1980RSPSB.207..187M}
{Marr}, D., \& {Hildreth}, E. 1980, Proceedings of the Royal Society of London
  Series B, 207, 187, \dodoi{10.1098/rspb.1980.0020}

\bibitem[{Mc~Dowell(2023)}]{leonumbers2023}
Mc~Dowell, J. 2023, Jonathan's Space Pages: Enormous (`Mega') Satellite
  Constellations.
\newblock \url{https://planet4589.org/space/con/conlist.html}

\bibitem[{McNally \& Rast(1999)}]{MCNALLY1999255}
McNally, D., \& Rast, R. 1999, Advances in Space Research, 23, 255,
  \dodoi{https://doi.org/10.1016/S0273-1177(99)00011-3}

\bibitem[{Mendes de Oliveira {et~al.}(2019)Mendes de Oliveira, Ribeiro,
  Schoenell, Kanaan, Overzier, Molino, Sampedro, Coelho, Barbosa, Cortesi,
  Costa-Duarte, Herpich, Hernandez-Jimenez, Placco, Xavier, Abramo, Saito,
  Chies-Santos, Ederoclite, Lopes de Oliveira, Gonçalves, Akras, Almeida,
  Almeida-Fernandes, Beers, Bonatto, Bonoli, Cypriano, Vinicius-Lima,
  de Souza, Fabiano de Souza, Ferrari, Gonçalves, Gonzalez,
  Gutiérrez-Soto, Hartmann, Jaffe, Kerber, Lima-Dias, Lopes,
  Menendez-Delmestre, Nakazono, Novais, Ortega-Minakata, Pereira, Perottoni,
  Queiroz, Reis, Santos, Santos-Silva, Santucci, Barbosa, Siffert, Sodré,
  Torres-Flores, Westera, Whitten, Alcaniz, Alonso-García, Alencar,
  Alvarez-Candal, Amram, Azanha, Barbá, Bernardinelli, Borges Fernandes,
  Branco, Brito-Silva, Buzzo, Caffer, Campillay, Cano, Carvano, Castejon,
  Cid Fernandes, Dantas, Daflon, Damke, de la Reza, de Melo de Azevedo,
  De Paula, Diem, Donnerstein, Dors, Dupke, Eikenberry, Escudero, Faifer,
  Farías, Fernandes, Fernandes, Fontes, Galarza, Hirata, Katena,
  Gregorio-Hetem, Hernández-Fernández, Izzo, Jaque Arancibia,
  Jatenco-Pereira, Jiménez-Teja, Kann, Krabbe, Labayru, Lazzaro, Lima Neto,
  Lopes, Magalhães, Makler, de Menezes, Miralda-Escudé, Monteiro-Oliveira,
  Montero-Dorta, Muñoz-Elgueta, Nemmen, Nilo Castellón, Oliveira, Ortíz,
  Pattaro, Pereira, Quint, Riguccini, Rocha Pinto, Rodrigues, Roig, Rossi,
  Saha, Santos, Schnorr Müller, Sesto, Silva, Smith Castelli, Teixeira,
  Telles, Thom de Souza, Thöne, Trevisan, de Ugarte Postigo,
  Urrutia-Viscarra, Veiga, Vika, Vitorelli, Werle, Werner, \& Zaritsky}]{SPLUS}
Mendes de Oliveira, C., Ribeiro, T., Schoenell, W., {et~al.} 2019, Monthly
  Notices of the Royal Astronomical Society, 489, 241,
  \dodoi{10.1093/mnras/stz1985}

\bibitem[{Nafornita \& Isar(2014)}]{7010797}
Nafornita, C., \& Isar, A. 2014, in 2014 11th International Symposium on
  Electronics and Telecommunications (ISETC), 1--4,
  \dodoi{10.1109/ISETC.2014.7010797}

\bibitem[{{Nogueira} {et~al.}(2012){Nogueira}, {Humberto Andrei}, {Da Silva
  Neto}, {Tang}, {Li}, {Mao}, {Penna}, {Teixeira}, {Yu}, \& {Fidencio
  Neto}}]{2012cosp...39.1381N}
{Nogueira}, E.~C., {Humberto Andrei}, A., {Da Silva Neto}, D., {et~al.} 2012,
  in 39th COSPAR Scientific Assembly, Vol.~39, 1381

\bibitem[{Occulus(2021)}]{camera}
Occulus. 2021, Occulus All-Sky Camera 180º.
\newblock \url{https://www.sxccd.com/product/oculus-all-sky-camera-180/}

\bibitem[{Pandeirada {et~al.}(2021)Pandeirada, Bergano, Neves, Marques,
  Barbosa, Coelho, \& Ribeiro}]{signals2010011}
Pandeirada, J., Bergano, M., Neves, J., {et~al.} 2021, Signals, 2, 122.
\newblock \url{https://www.mdpi.com/2624-6120/2/1/11}

\bibitem[{Peli(1990)}]{Peli:90}
Peli, E. 1990, J. Opt. Soc. Am. A, 7, 2032, \dodoi{10.1364/JOSAA.7.002032}

\bibitem[{{Perryman} {et~al.}(1997){Perryman}, {Lindegren}, {Kovalevsky},
  {Hog}, {Bastian}, {Bernacca}, {Creze}, {Donati}, {Grenon}, {Grewing}, {van
  Leeuwen}, {van der Marel}, {Mignard}, {Murray}, {Le Poole}, {Schrijver},
  {Turon}, {Arenou}, {Froeschle}, \& {Petersen}}]{Hipparcos}
{Perryman}, M.~A.~C., {Lindegren}, L., {Kovalevsky}, J., {et~al.} 1997,
  Astronomy and Astrophysics, 500, 501

\bibitem[{Pino~Gonçalves(2024)}]{CLOWNZenodo}
Pino~Gonçalves, L.~F. 2024, CLOWN - CLOud Watcher at Night,  Zenodo,
  \dodoi{10.5281/zenodo.10848202}

\bibitem[{{Pisano} {et~al.}(1998){Pisano}, {Zong}, B.M., {DeLuca}, {Johnston},
  {Muller}, {Braeuning}, \& {Pizer}}]{Pisano}
{Pisano}, E.~D., {Zong}, S., B.M., H., {et~al.} 1998, Journal of Digital
  Imaging, 11, 193, \dodoi{https://doi.org/10.1007/BF03178082}

\bibitem[{Pizer {et~al.}(1990)Pizer, Johnston, Ericksen, Yankaskas, \&
  Muller}]{109340}
Pizer, S., Johnston, R., Ericksen, J., Yankaskas, B., \& Muller, K. 1990, in
  [1990] Proceedings of the First Conference on Visualization in Biomedical
  Computing, 337--345, \dodoi{10.1109/VBC.1990.109340}

\bibitem[{Poelzl(2021)}]{AllSkEye}
Poelzl, M. 2021, AllSkEye.
\newblock \url{https://www.allskeye.com/}

\bibitem[{Prasad(2005)}]{PRASAD2005243}
Prasad, M. 2005, Space Policy, 21, 243,
  \dodoi{https://doi.org/10.1016/j.spacepol.2005.08.010}

\bibitem[{Rossi(2005)}]{Rossi}
Rossi, A. 2005, Serbian Astronomical Journal, 170, 1,
  \dodoi{10.2298/SAJ0570001R}

\bibitem[{Tingay {et~al.}(2013)Tingay, Kaplan, McKinley, Briggs, Wayth,
  Hurley-Walker, Kennewell, Smith, Zhang, Arcus, Bhat, Emrich, Herne,
  Kudryavtseva, Lynch, Ord, Waterson, Barnes, Bell, Gaensler, Lenc, Bernardi,
  Greenhill, Kasper, Bowman, Jacobs, Bunton, deSouza, Koenig, Pathikulangara,
  Stevens, Cappallo, Corey, Kincaid, Kratzenberg, Lonsdale, McWhirter, Rogers,
  Salah, Whitney, Deshpande, Prabu, Shankar, Srivani, Subrahmanyan, Ewall-Wice,
  Feng, Goeke, Morgan, Remillard, Williams, Hazelton, Morales,
  Johnston-Hollitt, Mitchell, Procopio, Riding, Webster, Wyithe, Oberoi, Roshi,
  Sault, \& Williams}]{Tingay_2013}
Tingay, S.~J., Kaplan, D.~L., McKinley, B., {et~al.} 2013, The Astronomical
  Journal, 146, 103, \dodoi{10.1088/0004-6256/146/4/103}

\bibitem[{Tonry(2013)}]{Pan-STARRS}
Tonry, J.~L. 2013, Philosophical Transactions of the Royal Society A:
  Mathematical, Physical and Engineering Sciences, 371, 20120269,
  \dodoi{10.1098/rsta.2012.0269}

\bibitem[{Tsai \& Yeh(2008)}]{4560077}
Tsai, C.-M., \& Yeh, Z.-M. 2008, IEEE Transactions on Consumer Electronics, 54,
  213, \dodoi{10.1109/TCE.2008.4560077}

\bibitem[{UCS(2022)}]{linksatelites}
UCS. 2022, Satellite Database.
\newblock \url{https://www.ucsusa.org/resources/satellite-database}

\bibitem[{UNOOSA(2022)}]{linkNacoesUnidas}
UNOOSA. 2022, United Nations Office for Outer Space Affairs.
\newblock \url{https://www.unoosa.org/oosa/osoindex/index.jspx?lf_id=}

\bibitem[{Vallado {et~al.}(2006)Vallado, Crawford, Hujsak, \&
  Kelso}]{inproceedings}
Vallado, D., Crawford, P., Hujsak, R., \& Kelso, T. 2006, Revisiting Spacetrack
  Report \#3, \dodoi{10.2514/6.2006-6753}

\bibitem[{van~der Walt {et~al.}(2014)van~der Walt, {S}ch\"onberger,
  {Nunez-Iglesias}, {B}oulogne, {W}arner, {Y}ager, {G}ouillart, {Y}u, \& the
  scikit-image contributors}]{scikit-image}
van~der Walt, S., {S}ch\"onberger, J.~L., {Nunez-Iglesias}, J., {et~al.} 2014,
  PeerJ, 2, e453, \dodoi{10.7717/peerj.453}

\bibitem[{{Williams} {et~al.}(2021){Williams}, {Hainaut}, {Otarola}, {Tan}, \&
  {Rotola}}]{2021Msngr.184....3W}
{Williams}, A., {Hainaut}, O., {Otarola}, A., {Tan}, G.~H., \& {Rotola}, G.
  2021, The Messenger, 184, 3, \dodoi{10.18727/0722-6691/5237}

\bibitem[{{Zhang} {et~al.}(2022){Zhang}, {Cai}, {Xue}, {Xue}, \&
  {Cai}}]{EstudoLEO}
{Zhang}, J., {Cai}, Y., {Xue}, C., {Xue}, Z., \& {Cai}, H. 2022, Space: Science
  and Technology, 2022, 9865174, \dodoi{10.34133/2022/9865174}

\bibitem[{Zhao(2010)}]{zhao2010linear}
Zhao, Y. 2010, Phd thesis, McMaster University, Ontario, Canada

\bibitem[{Zotti {et~al.}(2021)Zotti, Hoffmann, Wolf, Ch{\'e}reau, \&
  Ch{\'e}reau}]{Stellarium}
Zotti, G., Hoffmann, S.~M., Wolf, A., Ch{\'e}reau, F., \& Ch{\'e}reau, G. 2021,
  Journal of Skyscape Archaeology, 6, 221–258, \dodoi{10.1558/jsa.17822}

\end{thebibliography}
\bibliographystyle{aasjournal}



\end{document}